\documentclass[11pt]{article} 

\usepackage{times}
\usepackage{amsmath}
\usepackage{amsthm}
\usepackage{graphicx,epsfig}

\newtheorem{proposition}{Proposition}[section]

\newcommand {\E}{\mathbf E}
\newcommand {\T}{\mathcal T}

\renewcommand {\P}{\mathcal P}

\newcommand{\p}{\mathbf{p}}
\newtheorem{theorem}{Theorem}[section] 
\newtheorem{lemma}[theorem]{Lemma}

\newtheorem{claim}[theorem]{Claim}

\newtheorem{fact}[theorem]{Fact}
\newtheorem{define}{Definition}

\newtheorem{ex}{Example}[section]

\newcommand{\W}{\mathcal{W}}
\newcommand{\eat}[1]{}
\renewcommand{\paragraph}[1]{\medskip \noindent {\bf{#1}}}

\newcommand{\D}{\mathcal{D}}

\advance \topmargin by -\headheight
\advance \topmargin by -\headsep
\evensidemargin \oddsidemargin
\newcommand{\G}{\mathcal{G}}

\begin{document}
\title{Approximation Algorithms for Restless Bandit Problems\thanks{This paper combines and generalizes results presented in two  papers~\cite{GuhaM07c} and~\cite{GuhaMS08} that appeared in the FOCS '07 and SODA '09 conferences respectively.}}
\author{{Sudipto Guha\thanks{ Department of Computer and Information
 Sciences, University of Pennsylvania, Philadelphia PA 19104-6389.   Email: 
{\tt
 sudipto@cis.upenn.edu}. Research supported in part by an Alfred
 P. Sloan Research Fellowship, an NSF CAREER Award
CCF-0644119. } }\and {Kamesh Munagala\thanks{ Department of Computer 
Science,  Duke
 University,  Durham NC 27708-0129. Email: {\tt kamesh@cs.duke.edu}.  
Research
 supported by NSF via a CAREER award and grant CNS-0540347. }} \and Peng Shi\thanks{Duke 
University, Durham NC 27708. Email: {\tt peng.shi@duke.edu}. This research 
was supported by the Duke University Work-Study Program and by NSF 
award CNS-0540347.}} \date{} \maketitle

\begin{abstract}
 The {\em restless bandit} problem is
  one of the most well-studied generalizations of the celebrated
  stochastic multi-armed bandit problem in decision theory. In its
  ultimate generality, the restless bandit problem is known to be
  PSPACE-Hard to approximate to any non-trivial factor, and little
  progress has been made on this problem despite its significance in
  modeling activity allocation under uncertainty. 

  We consider a special case that  we call {\sc Feedback} MAB, where the reward obtained by playing
  each of $n$ independent arms varies according to an underlying on/off
  Markov process whose exact state is only revealed  when the
  arm is played. The goal is to design a policy for playing the arms in order
  to maximize the infinite horizon time average expected reward. This
  problem is also an instance of a Partially Observable Markov Decision Process
  (POMDP), and is widely studied in wireless scheduling
  and unmanned aerial vehicle (UAV) routing. Unlike the stochastic MAB problem, the {\sc Feedback}   MAB problem does not admit to greedy index-based optimal policies. The state of  the system at any time step encodes the beliefs about the states of
  different arms, and the policy decisions change these beliefs -- this   aspect complicates the design and analysis of simple algorithms.
  
  We develop a novel and fairly general duality-based algorithmic technique that yields a surprisingly simple and intuitive  $2+\epsilon$-approximate greedy policy to this problem. We then define a general sub-class of restless bandit problems that we term {\sc Monotone} bandits, for which our policy is a $2$-approximation. Our technique is robust enough to handle generalizations of these problems to incorporate various side-constraints such as blocking plays and switching costs. This technique is also of independent interest for other restless bandit problems, and we provide an example in non-preemptive machine replenishment. We finally show that our policies are closely related to the {\em Whittle index} that is  widely used for its simplicity and efficiency of computation. In fact, not only is our policy just as efficient to compute as the Whittle index, but in addition, it provides surprisingly strong constant factor guarantees even in cases where the Whittle index is provably polynomially worse.
  
 By presenting the first (and efficient) $O(1)$ approximations for non-trivial instances of restless bandits as well as   of POMDPs, our work initiates the study of approximation algorithms in   both these contexts.

\end{abstract}

\thispagestyle{empty}
\newpage
\setcounter{page}{1}

\newcommand{\Th}{\Theta}
\renewcommand{\S}{\mathcal{S}}
\renewcommand{\H}{\mathbf{T}}
\renewcommand{\T}{\mathbf{T}}
\renewcommand{\p}{\mathbf{p}}
\newcommand{\q}{\mathbf{q}}
\newcommand{\Eta}{\mathcal{E}}
\newcommand{\Z}{\mathcal{Z}}
\newcommand{\Y}{\mathcal{Y}}
\renewcommand{\L}{\mathcal{L}}
\renewcommand{\G}{\mathcal{G}}
\newcommand{\I}{\mathcal{I}}

\section{Introduction}
\label{sec:intro}
The celebrated multi-armed bandit problem (MAB) models the central
trade-off in decision theory between exploration and exploitation, or in
other words between learning about the state of a system and utilizing the
system. In this problem, there are $n$ competing options, referred to as
``arms,'' yielding unknown rewards $\{r_i\}$. Playing an arm yields a
reward drawn from an underlying distribution, and the information from the
reward observed partially resolves its distribution. The goal is to
sequentially play the arms in order to maximize reward obtained over some
time horizon.

Typically, the multi-armed bandit problem is studied under one of two
assumptions: 
\begin{enumerate}
\item The underlying reward distribution for each arm is fixed
but unknown, and a prior of this distribution is specified as input ({\em
  stochastic multi-armed bandits}~\cite{ABG49,book,Robbins,Wald47}); or
\item The underlying rewards can vary with time in an adversarial fashion,
and the comparison is against an optimal strategy that always plays one
arm, albeit with the benefit of hindsight ({\em adversarial multi-armed
  bandits}~\cite{Auer95,experts,FlaxmanKM05}).  
  \end{enumerate}Relaxing {\em both} the
assumptions simultaneously leads to the notorious {\em restless bandit}
problem in decision theory, which in its ultimate generality, is PSPACE
hard to even approximate~\cite{PapT}. In the last two decades, in spite of
the growth of approximation algorithms and the numerous applications of
restless bandits~\cite{glaze1,glaze2,glaze3,glaze4,
  zhao2,leny,nino2,weiss2,whittle2}, the approximability of these have remained
unexplored.  In this paper, we provide a general algorithmic technique that yields the first $O(1)$ approximations to a large class of these problems that are commonly studied in practice.

\paragraph{}
An important subclass of restless bandit problems are situations where the
system is agnostic of the exploration -- or the exploration gives us
information about the state of the system but does not interfere with the
evolution of the system. One such problem is the {\sc Feedback} MAB which
models {\em opportunistic multi-channel access} at a wireless
node~\cite{GuhaM07c,javidi2,zhao07}: The bandit corresponds to a wireless
node with access to multiple noisy channels (arms). The state of the arm
is the state (good/bad) of the channel, which varies according to a bursty
$2$-state Markov process.  Playing the arm corresponds to transmitting on
the channel, yielding reward if the transmission is successful (good
channel state), and at the same time revealing to the transmitter the
current state of the channel. This corresponds to the Gilbert-Elliot
model~\cite{gilbert} of channel evolution. The goal is to find a
transmission policy of choosing one channel to transmit on every time
step, that maximizes the long-term transmission rate. {\sc Feedback MAB}
also models Unmanned Aerial Vehicle (UAV) routing~\cite{leny}: the arms
are locations of possibly interesting events, and whether a location is
interesting or uninteresting follows a $2$-state Markov processes.
Visiting a location by the UAV corresponds to playing the arm, and yields
reward if an interesting event is detected. The goal is to find a routing
policy that maximizes the long-term average reward from interesting
events.

This problem is also a special case of {\em Partially Observable Markov
  Decision Processes} or POMDPs~\cite{kaelbling,pomdp1,pomdp2}. The state
of each arm evolves according to a Markov chain whose state is only
observed when the arm is played. The player's partial information,
encapsulated by the last observed state and the number of steps since last
playing, yields a belief on the current state. (This belief is simply a
probability distribution for the arm being good or bad.) The player uses
this partial information in making the decision about which arm to play
next, which in turn affects the information at future times. While such
POMDPs are widely used in control theory, they are in general notoriously
intractable~\cite{book,kaelbling}.  In this paper we provide the first
$O(1)$ approximation for the {\sc Feedback} MAB and a number of its
important extensions.  This represents the first approximation guarantee
for a  POMDP, and the
first guarantee for a MAB problem with time-varying rewards that compares
to an optimal solution allowed to switch arms at will.

\medskip Before we present the problem statements formally, we survey
literature on the stochastic multi-armed bandit problem. (We discuss
adversarial MAB after we present our model and results.)

\subsection{Background: Stochastic MAB and Restless Bandits}

The stochastic MAB  was first formulated by Arrow
{\em et al}~\cite{ABG49} and Robbins~\cite{Robbins}. It resides under a
Bayesian (or decision theoretic) setting: we successively choose between
several options given some prior information (specified by distributions),
and our beliefs are updated via Bayes' rule conditioned on the results of
our choices (observed rewards). 

More formally, we are given a ``bandit"
with $n$ independent arms.  Each arm $i$ can be in one of several states
belonging to the set $\S_i$. At any time step, the player can play one
arm. If arm $i$ in state $k \in \S_i$ is played, it transitions in a
Markovian fashion to state $j \in \S_i$ w.p. $q^i_{kj}$, and yields reward
$r^i_{k} \ge 0$. The states of arms that are not played stay the same.
The initial state models the prior knowledge about the arm. The states in
general capture the posterior conditioned on the observations from
sequential plays. The problems is, given the initial states of the arms,
find a {\em policy} for playing the arms in order to maximize one of the
following infinite horizon quantities: $\sum_{t=0}^{\infty} R_t \beta^t$
(discounted reward), or $\lim_{t \rightarrow \infty} \frac1t
\sum_{t=0}^{\infty} R_t$ (average reward), where $R_t$ is the expected
reward of the policy at time step $t$ and $\beta \in (0,1)$ is a discount
factor. A {\em policy} is a (possibly implicit) specification of {\em fixing
  up front} which arm (or distribution over arms) to play for every
possible joint state of the arms. 

It is well-known that Bellman's equations~\cite{book} yield the optimal
policy by dynamic programming. The main issue in the stochastic setting is
in {\em efficiently computing and succinctly specifying} the optimal
policy: The input to an algorithm specifies the rewards and transition
probabilities for each arm, and thus has size linear in $n$, but the state
space is exponential in $n$.  We seek polynomial-time algorithms (in terms
of the input size) that compute (near-) optimal policies with poly-size
specifications. Moreover, we require the policies to be executable each
step in poly-time.

Note that since a policy is a fixed (possibly randomized) mapping from the
{\em exponential size} joint state space to a set of actions, ensuring
poly-time computation and execution often requires simplifying the
description of the optimal policy using the problem structure. The
stochastic MAB problem is the most well-known decision problem for which
such a structure is known: The {\em optimal} policy is a greedy policy
termed the {\sc Gittins} index policy~\cite{GJ74,tsitsiklis,book}.  In
general, an index policy specifies a single number called ``index" for
each state $k \in \S_i$ for each arm $i$, and at every time step, plays
the arm whose current state has the highest index. Index policies are
desirable since they can be compactly represented, so they are the heuristic
method of choice for several MDP problems. In addition, index policies are
also optimal for several generalizations of the stochastic MAB, such as
arm-acquiring bandits~\cite{whittle} and branching bandits~\cite{weiss}.
In fact, a general characterization of problems for which index policies
are optimal is now known~\cite{BV06}.

\paragraph {Restless Bandits.} In the stochastic MAB problem, the
underlying reward distributions for each arm are fixed but unknown.
However, if the rewards can vary with time, the problem stops admitting
optimal index policies or efficient solutions. The problem now needs to be
modeled as a {\em restless bandit} problem, first proposed by
Whittle~\cite{whittle2}. The problem statement of the restless bandits is
similar to stochastic MAB, except that when arm $i$ in state $k \in \S_i$
is not played, it's state evolves to $j \in \S_i$ with probability
$\tilde{q}^i_{kj}$. Therefore, the state of each arm varies according to
an {\em active} transition matrix $q$ when the arm is played, and
according to a {\em passive} transition matrix $\tilde{q}$ if the arm is
not played. Unlike the stochastic MAB problem, which is interesting only
in the discounted reward setting\footnote{Playing the arm with the highest
  long-term average reward exclusively is the trivial optimal policy for
  stochastic MAB in the infinite-horizon average reward setting.}, the
restless bandit problem is interesting even in the {\em infinite horizon
  average reward} setting -- this is the setting in which the problem has
been typically studied, and so we limit ourselves to this setting in this
paper.  It is relatively straightforward to show that no index policy can
be optimal for these problems; in fact, Papadimitriou and
Tsitsiklis~\cite{PapT} show that for $n$ arms, even when all $q$ and
$\tilde{q}$ values are either $0$ or $1$ (deterministic transitions),
computing the optimal policy is a PSPACE-hard problem. Their proof in fact
shows that deciding if the optimal reward is non-zero is also PSPACE-hard,
hence ruling out any approximation algorithm as well.

On the positive side, Whittle~\cite{whittle2} presents a poly-size LP
relaxation of the problem. In this relaxation, the constraint that exactly
one arm is played per time step is replaced by the constraint that one arm
{\em on average} is played per time step. In the LP, this is the only
constraint connecting the arms. (Such decision problems have been termed
{\em weakly coupled} systems~\cite{hawkins,Adelman}.)  Based on the
Lagrangean of this relaxation, Whittle~\cite{whittle2} defines a heuristic
index that generalizes the Gittins index. This is termed the {\em Whittle
  Index} (see Section~\ref{sec:index}). Though this index is widely used
in practice and has excellent empirical
performance~\cite{glaze1,glaze2,glaze3,glaze4,zhao2,leny,weiss2}, the
known theoretical guarantees (~\cite{weiss2,glaze2}) are very weak.
In summary, despite being very well-motivated and extensively studied,
there are almost no positive results on approximation guarantees for the
restless bandit problems.

\subsection{Results and Roadmap} 
 We provide the first approximation algorithm for both a restless
  bandit problem and a partially observable Markov decision problem by
  providing a $2+\epsilon$-approximate index policy for the {\sc Feedback} MAB problem
  which belongs to both classes. We show several other results; however, before presenting the specifics, we place our contribution in the context of existing techniques in control theory.
  
\paragraph{Technical Contributions.} Our algorithmic technique for this problem (Section~\ref{app:feedback}) involves solving (in polynomial time) the Lagrangean of Whittle's LP relaxation for a suitable (and subtle) ``balanced" choice of the Lagrange multiplier, converting this into a feasible index policy, and using an amortized accounting of the reward for the analysis. We show that this technique is closely related to the Whittle index~\cite{whittle2,leny,javidi2}, and in fact, provide the first approximation analysis of (a subtle variant of) the Whittle index which is widely used in control theory literature in  the context of {\sc Feedback} MAB problems (Section~\ref{sec:index}). We believe that analyzing  the performance guarantees of the numerous indices used in the  literature will increase and our analysis will provide an useful  template.

However, the key difference between Whittle's index and our index policy is the following: The former chooses one Lagrange multiplier  (or index) per state of each arm, with the policy playing the arm with the largest index. This has the advantage of separate efficient computations for different arms; and in addition, such a policy (the Gittins index policy~\cite{GJ74}) is known to be optimal for the stochastic MAB. However, it is well-known~\cite{asawa,BS94,ML} that this intuition about playing the arm with  the largest index being optimal becomes increasingly invalid when complicated side-constraints such as time-varying rewards ({\sc Feedback MAB}), blocking plays, and switching costs are introduced. In fact, we show a concrete problem in Section~\ref{sec:beyond} where  the Whittle index has a $\Omega(n)$ performance gap. 

In contrast to the Whittle index, our technique chooses a {\em single global} Lagrange multiplier via a careful accounting of the reward, and develops a feasible policy from it. Unlike the Whittle index, this technique is sufficiently robust to encompass a large number of often-used variants of {\sc  Feedback} MAB problems: Plays with varying duration (Section~\ref{app:vary}), switching costs (Section~\ref{app:switch}), and observation costs (Section~\ref{app:probe}). In fact, we identify a general {\sc Monotone} condition in restless bandit problems under which our technique applies (Section~\ref{sec:recover}).  Furthermore, our technique provides $O(1)$ approximations to other classic restless bandit problems even when Whittle's index is polynomially sub-optimal: We show an example in the non-preemptive machine replenishment problem (Section~\ref{sec:beyond}). Finally, since our technique is based on solving the Lagrangean\footnote{This aspect is explicit in Sections~\ref{app:feedback} and~\ref{sec:index}. However, in Sections~\ref{sec:recover}--\ref{sec:beyond}, we have presented our algorithm in terms of first solving a linear program. However, it is easy to see that this is equivalent to solving the Lagrangean, and hence to the computation required for Whittle's index. The details are quite standard and can be reconstructed from those in Sections~\ref{app:feedback} and~\ref{sec:index}.} (just like the Whittle index), the computation time is comparable to that for such indices.

In summary, our technique succeeds in finding the first provably approximate policies for widely-studied control problems, {\em without sacrificing efficiency} in the process. We believe that the generality of this technique will be useful for exploring other useful variations of  these problems as well as providing an alternate algorithm for practitioners.

\paragraph{Specific Results.} In terms of specific results, the paper is organized as follows:
\begin{itemize}
\item We begin by presenting a $2+\epsilon$-approximation for {\sc
    Feedback} bandits in Section~\ref{app:feedback}.  We also provide a $e/(e-1)$ integrality gap
  instance showing that out analysis is nearly tight.  
\item In Section~\ref{sec:index} we show that our analysis technique can
  be used to prove that a thresholded variant of the Whittle index is a
  $2$ approximation. We also show instances where the reward of any index
  policy is at least $1+\Omega(1)$ factor from the reward of the optimal
  policy.  Therefore although the Whittle index is not optimal, our result sheds light on its observed
  superior performance  in this specific context. 
\item In Section~\ref{sec:recover} we generalize the result in Section~\ref{app:feedback} to define a general sub-class of restless bandit problems based on a critical set of
  properties: Separability and monotonicity. For this subclass, termed {\sc Monotone} bandits (which generalizes {\sc Feedback MAB}), we provide a $2$ approximation by generalizing the technique
  in Section~\ref{app:feedback}.  Our technique now introduces a balance constraint in the dual of the natural LP relaxation, and constructs the index policy from the optimal dual solution. We further show that in the absence  of monotonicity or separability,  the 
  problem is either NP-Hard to approximate, or has unbounded integrality gap respectively.
\item In Section~\ref{app:vary} we extend {\sc Feedback} MAB (as well as
  {\sc Monotone} bandits) to consider multiple simultaneous blocking plays of
  varying durations.
\item In Section~\ref{app:switch} we extend {\sc Feedback} MAB (and
  {\sc Monotone} bandits) to consider switching costs. 
\item In Section~\ref{app:probe} we extend {\sc Feedback} MAB to a variant
  where the information acquisition is varied, namely, an arm has to be
  explicitly probed at some cost to obtain its state.
\item In Section~\ref{sec:beyond}, we derive a $2$-approximation for a
  classic, restless bandit problem called non-preemptive machine
  replenishment~\cite{book,trivedi,Shi2007}. We also show that the Whittle
  Index for this problem has a $\Omega(n)$ factor worse performance
  compared to the optimal policy. Thus the technique introduced in this
  paper can be superior to Whittle index or similar policies.
\end{itemize}

\subsection{Related Work} 
\noindent {\bf Contrast with the Adversarial MAB Problem.} 
While our problem formulations are based on the stochastic MAB problem, one might be interested in a
formulation based on the adversarial MAB~\cite{Auer95,Lai}. Such a formulation might be to assume
that rewards can vary adversarially, and that the objective is to compete
with a restricted optimal solution that always plays the same arm but with
the benefit of hindsight.

These different formulations result in fundamentally different problems.
Under our formulation, the difficulty is computational: we want to compute
{\em policies} for playing the arms, assuming stochastic models of how the
system varies with time. under the adversarial formulation, the difficulty
is {\em informational}: we would be interested in the {\em regret} of not
having the benefit of hindsight.  A sequence of papers show near-tight
regret bounds in fairly general
settings~\cite{auer2,auer3,Auer95,experts,FlaxmanKM05,Lai,LittlestoneW}.
However, applying this framework is not satisfying: It is straightforward
to show that a policy for {\sc Feedback MAB} that is allowed to switch
arms can be $\Omega(n)$ times better than a policy that is not allowed to
do so (even assuming hindsight). Another approach would be to define each
policy as an ``expert", and use the low-regret experts
algorithm~\cite{experts}; however, the number of policies is
super-exponentially large, which would lead to weak regret bounds, along
with exponential-size policy descriptions and exponential per-step
execution time.

We note that developing regret bounds in the presence of changing
environments has received significant interest recently in computational
learning~\cite{auer2,Auer95,defarias,kakade,slivkins}; however, this
direction requires strong assumptions such as bounded switching between
arms~\cite{auer2} and slowly varying environments~\cite{kakade,slivkins},
both of which assumptions are inapplicable to {\sc Feedback MAB}.  In
 independent work, Slivkins and Upfal~\cite{slivkins}
consider the modification of {\sc Feedback MAB} where the underlying state
of the arms vary according to a reflected Brownian motion with bounded
variance.  As discussed in \cite{slivkins}, this problem is technically
very different from ours, even requiring different performance metrics.

\paragraph{Other Related Work.} 
The results in \cite{GuhaM07b,GoelGM06,GuhaM06,GuhaMS06} consider variants
of the stochastic MAB where the underlying reward distribution does not
change and only a limited time is allotted to learning about this
environment. Although several of these results use LP rounding, they have
little connection to the duality based framework considered here.

Our duality based framework shows a $2$-approximate index policy for
non-preemptive machine replenishment (Section~\ref{sec:beyond}). Elsewhere,
Munagala and Shi~\cite{Shi2007} considered the special case of preemptive
machine replenishment problem, for which the Whittle index is equivalent
to a simple greedy scheme. They show that this greedy policy, though not
optimal, is a $1.51$ approximation. However, the techniques there are
based on queuing analysis, and do not extend to the non-preemptive case
where the Whittle index can be an arbitrarily poor approximation (as shown
in Section~\ref{sec:beyond}).

Our solution technique differs from primal-dual approximation
algorithms~\cite{vazirani01} and online algorithms~\cite{young}, which
relax either the primal or the dual complementary slackness conditions
using a careful dual-growing procedure.  Our index policy and associated
potential function analysis crucially exploit the structure of the {\em
  optimal} dual solution that is gleaned using both the {\em exact} primal
as well as dual complementary slackness conditions. Furthermore, our
notion of dual balancing is very different from that used by Levi {\em et
  al}~\cite{levi} for designing online algorithms for stochastic inventory
management.

\section{The {\sc Feedback} MAB Problem}
\label{app:feedback}
In this problem, first formulated independently
in~\cite{GuhaM07c,zhao07,javidi2,leny}, there is a bandit with $n$
independent arms. Arm $i$ has two states: The good state $g_i$ yields
reward $r_i$, and the bad state $b_i$ yields no reward. The evolution of
state of the arm follows a bursty $2$-state Markov process which does not
depend on whether the arm is played or not at a time slot. Let $s_{it}$
denote the state of arm $i$ at time $t$. Denote the transition
probabilities of the Markov chain as follows: $\Pr[s_{i(t+1)} = g_i |
s_{it} = b_i] = \alpha_i$ and $\Pr[s_{i(t+1)} = b_i | s_{it} = g_i] =
\beta_i$. The $\alpha_i, \beta_i, r_i$ values are specified as input. The
``burstiness'' assumption simply means $\alpha_i + \beta_i \le 1-\delta$
for some small $\delta > 0$ specified as part of the input.  The evolution
of states for different arms are independent.  Any policy chooses at most
one arm to play every time slot. Each play is of unit duration, yields
reward depending on the state of the arm, and reveals to the
policy the current state of that arm. When an arm is not played, the true
underlying state cannot be observed, which makes the problem a POMDP.  The
goal is to find a policy to play the arms in order to maximize the
infinite horizon average reward.

\medskip
First observe that we can change the reward structure of {\sc Feedback} MAB so that
when an arm is played, we obtain reward from the {\em last-observed} state
instead of the currently observed state. This does not change the average
reward of any policy. This allows us to encode all the state of each arm as follows.

\begin{proposition}
From the perspective of any policy, the state of any arm 
can be encoded as $(s,t)$, which denotes that it was last observed $t \ge
1$ steps ago to be in state $s \in \{g_i,b_i\}$.
\end{proposition}

Note that any policy maps each possible joint state of $n$ arms into an action of which arm to play. Such a mapping has size exponential in $n$. The standard heuristic is to consider index policies: Policies which define an ``index"  or number for each state $(s_i,t)$ and play the arm with the highest current index. The following theorem shows that playing the arm with the highest myopic reward does not work, and that index policies in general are non-optimal. Therefore, our problem is interesting and the best we can hope for with index policies is a $O(1)$ approximation.

\begin{theorem}
\label{thm:no-opt-index} (Proved in Appendix~\ref{app:omitted})
  For {\sc Feedback MAB}, the reward of the optimal policy has
  an $\Omega(n)$ gap against that of the myopic index policy and an $\Omega(1)$ gap against that of the optimal index policy.
\end{theorem}

\paragraph{Roadmap.}
In this section, we show that a simple index policy is a
$(2+\epsilon)$ approximation. This is based on a natural LP relaxation
suggested by Whittle which we discuss in Section~\ref{sec:whittleLP}; this formulation will have infinitely many constraints.  We then consider the Lagrangean of this formulation in Section~\ref{app:gap2}, and analyze its structure via duality, which enables computing its optimal solution in polynomial time. At this point, we deviate significantly from previous literature, and present our main contribution in Section~\ref{sec:bala}: A subtle and powerful ``balanced" choice of the Lagrange multiplier, which enables the design of an intuitive index policy,  {\sc BalancedIndex}, along with an equally intuitive analysis. We use duality and potential function arguments to show that the policy is $(2+\epsilon)$ approximation.
We conclude by showing that the gap of
Whittle's relaxation is $e/(e-1) \approx 1.58$, indicating that our
analysis is reasonably tight. This analysis technique generalizes easily
(explored in Sections~\ref{sec:recover} -- \ref{sec:beyond}) and has rich connections to other index
policies, most notably the Whittle index (explored in Section~\ref{sec:index}).

\subsection{Whittle's LP}
\label{sec:whittleLP}
Whittle's LP  is obtained by effectively replacing the hard constraint of
playing one arm per time step, with allowing multiple plays per step but requiring one play per step {\em on average}. Hence, the LP is a {\em relaxation} of the optimal policy.

\begin{define}
\label{def:v} 
Let $v_{it}$ be the probability of the arm $i$ being in
state $g_i$ when it was last observed in state $b_i$ exactly $t$ steps
ago. Let $u_{it}$ be the same probability when the last
observed state was $g_i$. We have:
$$v_{it} = \frac{\alpha_i}{\alpha_i+\beta_i} (1- (1-\alpha_i- \beta_i)^t) \qquad \mbox{and} \qquad u_{it} = \frac{\alpha_i}{\alpha_i+\beta_i} + \frac{\beta_i}{\alpha_i+\beta_i}  (1-\alpha_i- \beta_i)^t$$
\end{define}

\begin{fact}
\label{fact1}
The functions  $v_{it}$ and $1-u_{it}$ are monotonically increasing and concave functions of $t$.
\end{fact}

We now present Whittle's LP, and interpret it in the lemma that immediately follows. 

{\small 
\[ \mbox{Maximize       } \ \ \ \  \sum_{i=1}^n \sum_{t \ge 1} r_i x^i_{gt} \qquad   \] 
\[ \begin{array}{rcll}
\sum_{i=1}^n \sum_{s \in \{g,b\}} \sum_{t \ge 1} x^i_{st} & \le
& 1 & \\
\sum_{t \ge 1} \sum_{s \in \{g,b\}} (x^i_{st} + y^i_{st}) & \le
& 1 & \\
x^{i}_{st} + y^i_{st} & = & y^i_{s(t-1)} & \forall i, s \in \{g,b\}, t\ge 2\\  
x^i_{g1} + y^i_{g1}& = & \sum_{t\geq 1} x^i_{bt} v_{it} + \sum_{t\geq 1} u_{it} x^i_{gt} 
 & \forall i\\
x^i_{b1} + y^i_{b1} & = & \sum_{t\geq 1} x^i_{bt} (1- v_{it}) + \sum_{t\geq 1} (1-u_{it}) x^i_{gt} 
 & \forall i\\
y^i_{st},x^i_{st} & \ge & 0 & \forall i, s \in \{g,b\}, t\end{array}\]
}

\begin{lemma}
\label{lem:optopt}
The optimal objective to Whittle's LP, $OPT$, is at least the value of the optimal policy.
\end{lemma}
\begin{proof}
Consider the optimal policy. In the execution of this policy, for each arm $i$ and state $(s,t)$ for $s \in \{g,b\}$, let the variable $x^i_{st}$
denote the probability (or fraction of time steps) of the event: Arm $i$ is in state $(s,t)$ and gets played. Let $y^i_{st}$ correspond to the probability of the event that the state is $(s,t)$ and the arm is not played. Since the underlying Markov chains are ergodic, the optimal policy when executed is ergodic, and the above probabilities are well-defined. 

Now, at any time step, some arm $i$ in state $(s,t)$ is played, which implies the $x^i_{st}$ values are probabilities of mutually exclusive events. This implies they satisfy the first constraint in the LP. Similarly, for each arm $i$, at any step,  this arm is in some state $(s,t)$ and is either played or not played, so that  the $x^i_{st}, y^i_{st}$ correspond to mutually exclusive events. This implies that for each $i$, they satisfy the second constraint. For any arm $i$ and state $(s,t)$, the LHS of the third constraint is the probability of being in this state, while the RHS is the probability of entering this state; these are clearly identical in the steady state. For arm $i$, the LHS of the fourth (resp. fifth) constraint is the probability of being in state $(g,1)$ (resp. $(b,1)$), and the RHS is the probability of entering this state; again, these are identical.

This shows that the probability values defined for the execution of the optimal policy are feasible for the constraints of the LP. The value of the optimal policy is precisely $ \sum_{i=1}^n \sum_{t \ge 1} r_i x^i_{gt} $, which is at most $OPT$ -- the maximum possible objective for the LP. 
\end{proof}

The above LP encodes in one variable $x^i_{st}$ the probability the arm $i$ is in state $(s,t)$ and gets played; however, we note that in the optimal policy, this decision to play actually depends on the joint state of all arms. This separation of the joint probabilities into individual probabilities effectively relaxes the condition of having one play per step, to allowing multiple plays per step but requiring one play per step {\em on average}. While the policy generated by Whittle's LP is infeasible, the relaxation allows us to compute an {\em upper-bound} on the value of the optimal feasible policy. 

We note $y^i_{st} =\sum_{t'>t} x^i_{st'}$. It is convenient to eliminate the variables $y^i_{st}$ by
substitution and the last two constraints collapse into the same constraint.
Thus, we have the natural LP formulation shown in
Figure~\ref{fig:basic}. We note that the first constraint can either be an inequality ($\le$) or an equality; w.l.o.g., we use equality, since we can add a dummy arm that does not yield any reward on playing. 

\begin{figure*}[htbp]
\centerline{\framebox{\small
\begin{minipage}{6in}
\[ \mbox{Maximize       } \ \ \ \  \sum_{i=1}^n \sum_{t \ge 1} r_i x^i_{gt} \qquad \qquad \mbox{\sc (Whittle)}  \] 
\[ \begin{array}{rcll}
\sum_{i=1}^n \sum_{s \in \{g,b\}} \sum_{t \ge 1} x^i_{st} & = & 1 & \\
\sum_{t \ge 1} \sum_{s \in \{g,b\}} t x^i_{st} & \le & 1 & \forall i \\
\sum_{t \ge 1} x^i_{bt} v_{it} & = &  \sum_{t\ge 1} x^i_{gt} (1-u_{it})&
\forall i\\
x^i_{st} & \ge & 0 & \forall i, s\in\{g,b\}, t\end{array}\]
\end{minipage}}
}
\caption{\label{fig:basic}The linear program {\sc (Whittle)} for the {\sc
    feedback} MAB problem.}
\end{figure*}

\medskip
From now on, let $OPT$ denote the value of the optimal solution to {\sc (Whittle)}. The LP in its current form has infinitely many constraints; we will now show that this LP can be solved in polynomial time to arbitrary precision by finding structure in the Lagrangean.

\subsection{Decoupling Arms via the Lagrangean}
\label{app:gap2}
 In {\sc (Whittle)}, the only constraint connecting  different arms is the constraint:  
{\small $$\sum_{i=1}^n \sum_{s \in \{g,b\}} \sum_{t \ge 1} x^i_{st} = 1 $$}
 We absorb this constraint into the objective via
Lagrange multiplier $\lambda \ge 0$ to obtain the following 
objective:
{\small $$\mbox{Max.}\ \ \ \lambda + G(\lambda) \ \ \equiv \ \  \lambda + \sum_{i=1}^n\sum_{t \ge 1} \left(r_i x^i_{gt} - \lambda \left(x^i_{gt} + x^i_{bt} \right) \right) \qquad \qquad \left(\mbox{ {\sc LPLagrange}}(\lambda)\right)$$
\[ \begin{array}{rcll}
\sum_{t \ge 1} \sum_{s \in \{g,b\}} t x^i_{st} & \le & 1 & \forall i \\
\sum_{t \ge 1} x^i_{bt} v_{it} & = &  \sum_{t\ge 1} x^i_{gt} (1-u_{it})&
\forall i\\
x^i_{st} & \ge & 0 & \forall i, s\in\{g,b\}, t\end{array}\]
 }

 Through the Lagrangean, we have effectively removed the only constraint that connected multiple arms. 
{\sc LPLagrange}$(\lambda)$ now yields $n$ {\em disjoint} maximization problems,
one for each arm $i$: At any time step, arm $i$ can be played
(and reward obtained from it), or not played. Whenever the arm is played,
we incur a penalty $\lambda$ in addition to the reward. The goal is to maximize the expected reward minus cost.  Note that if the penalty is zero, the arm is played every step, and if the penalty is sufficiently large, the optimal
solution would be to never play the arm. 

\begin{define}
For each arm $i$, let $L_i(\lambda)$ denote the optimal policy, and let $H_i(\lambda)$ denote the optimal reward minus penalty. Note that the global reward minus penalty is the sum for each arm: $G(\lambda) = \sum_{i=1}^n H_i(\lambda)$.
\end{define}

\subsubsection{Characterizing the Optimal Single-arm Policy} 
We first show that the optimal policy $L_i(\lambda)$ for any arm $i$ belongs to the class of policies $\P_i(t)$ for $t \ge 1$, whose
specification is presented in Figure~\ref{fig:pk}. Intuitively, step (1) corresponds to exploitation, and step (2) to
exploration. Set $\P_i(\infty)$ to be the policy that never plays the arm.

\begin{figure*}[htbp]
\framebox{\small
\begin{minipage}{6.0in}
{\bf Policy $\P_i(t)$:}
\begin{tabbing}
1.\ \ \= If the arm was just observed to be in state $g$, then play the arm. \\
2.\> If the arm was just observed to be in state $b$, wait $t-1$ steps and play the arm.
\end{tabbing}
\end{minipage}
}
\caption{\label{fig:pk} The Policy $\P_i(t)$.}
\end{figure*}

To show this, we take with the dual of {\sc LPLagrange}$(\lambda)$:

{\small
\[ \mbox{Minimize } \ \ \ \ \lambda + \sum_{i=1}^n h_i \qquad
\qquad\qquad\qquad \mbox{\sc Whittle-Dual}(\lambda) \]
\[\begin{array}{rcll}
  \lambda + t h_i & \ge & r_i   - (1-u_{it}) p_i  & \forall i,t \ge 1 \\
  \lambda + t h_i & \ge &  v_{it} p_i  & \forall i,t \ge 1\\
 h_i & \ge & 0 & \forall i
\end{array} \]}

The fact that the optimal single arm policy $L_i(\lambda)$ belongs to the class $\{ P_i(t)\}$ comes from (5) of the following lemma.

\begin{lemma}
\label{lem:approxbalance}
For any $\lambda \ge 0$,  in the optimal solution to {\sc Whittle-Dual}$(\lambda)$, for any arm with $h_i > 0$: 
\begin{enumerate}
\item $h_i = H_i(\lambda)$.
\item $p_i \ge 0$.
\item For some $t_i \ge 1$, $x^i_{bt_i} > 0$ and $\lambda + t_i h_i =  v_{it_i} p_i$.
\item $x^i_{g1} > 0$ and $\lambda + h_i  = r_i   - \beta p_i$.
\item The optimal single-arm policy for arm $i$ is $L_i(\lambda)  = \P_i(t_i)$.
\end{enumerate}
\end{lemma}
\begin{proof}
The first part follows the definition of strong duality. The problem {\sc LPLagrange}$(\lambda)$, ignoring the constant $\lambda$ in the objective, separates into $n$ separate LPs, one for each arm. The dual objective for arm $i$ is precisely $h_i$, which must be the same as the primal objective, $H_i(\lambda)$.

If $h_i = H_i(\lambda) > 0$, the solution to the LP for arm $i$ is the policy $L_i(\lambda)$. In order to have non-zero $H_i(\lambda)$, such a policy must play the arm first in some state $(b,t_i)$ and state $(g,t'_i)$. Since $x^i_{st}$ is the probability this policy plays in state $(s,t)$, this implies $x^i_{bt_i} > 0$ and $x^i_{gt'_i} > 0$.

Since $x^i_{bt_i} > 0$, by complementary slackness, we have $\lambda + t_i h_i =  v_{it_i} p_i$. Since the LHS is at least zero, this implies $p_i \ge 0$. This proves parts (2) and (3).

To see part (4), observe that for the set of constraints  $\lambda + t h_i  \ge  r_i   - (1-u_{it}) p_i$, since $1 - u_{it} $ is a monotonically increasing function of $t$, the RHS is monotonically decreasing in $t$. Since the LHS is monotonically increasing, if the LHS and RHS are equal, they have to be so for $t = 1$. Now, since  $x^i_{gt'_i} > 0$, by complementary slackness, $\lambda + t'_i h_i  =  r_i   - (1-u_{it'_i}) p_i$. By the above argument, $t'_i = 1$, which completes the proof of part (4).

Since $x^i_{g1} > 0$ and $x^i_{bt_i} > 0$, the optimal policy for $L_i(\lambda)$ plays the arm in state $(g,1)$ and in state $(b,t_i)$, which is precisely the description of  $\P_i(t_i)$. This proves part (5).
\end{proof}

It will be instructive to interpret the problem $L_i(\lambda)$ as follows: Amortize the reward so that for each play, the arm $i$ yields a steady reward of $\lambda$. The goal is to find the single-arm policy that optimizes the {\em excess} reward per step over and above the amortized reward $\lambda$ per play.  As we have shown above, the optimal value for this problem is precisely $H_i(\lambda)$, and the policy $L_i(\lambda)$ that achieves this belongs to the class $ \{\P_i(t), t \ge 1\}$.

\subsubsection{Solving {\sc LPLagrange}$(\lambda)$}
Having decomposed the program {\sc LPLagrange}$(\lambda)$ into independent maximization problems for each arm, and having characterized the optimal single-arm policies, we can now solve the program in polynomial time. It will turn out this can be solved by  simple function maximization via closed form expressions.

\begin{define}
\label{def:rq}
For policy $\P_i(t)$, let $R_i(t)$ denote the expected per-step reward, and let $Q_i(t)$ denote the expected rate of play.  Let $F_i(\lambda,t) = R_i(t) - \lambda Q_i(t)$ denote the value of $\P_i(t)$. Also define:
\begin{equation}
\label{t}t_i(\lambda) = \mbox{argmax}_{t \ge 1} F_i(\lambda,t) = \mbox{argmax}_{t \ge 1} R_i(t) - \lambda Q_i(t)
\end{equation}
Finally, let $R_i(\lambda) = R_i(t_i(\lambda))$ and $Q_i(\lambda) = Q_i(t_i(\lambda))$
\end{define}
 
Note that the optimal reward minus cost for arm $i$ is simply $H_i(\lambda)=\max_{t \ge 1} R_i(t) - \lambda Q_i(t)=R_i(t_i)-\lambda Q_i(t_i)$. Since each $\P_i(t)$ corresponds to a Markov Chain, it is straightforward to obtain closed form expressions for $R_i(t)$ and $Q_i(t)$.
 
\begin{figure}[htbp]
\centerline{\includegraphics[angle=0,width=3in]{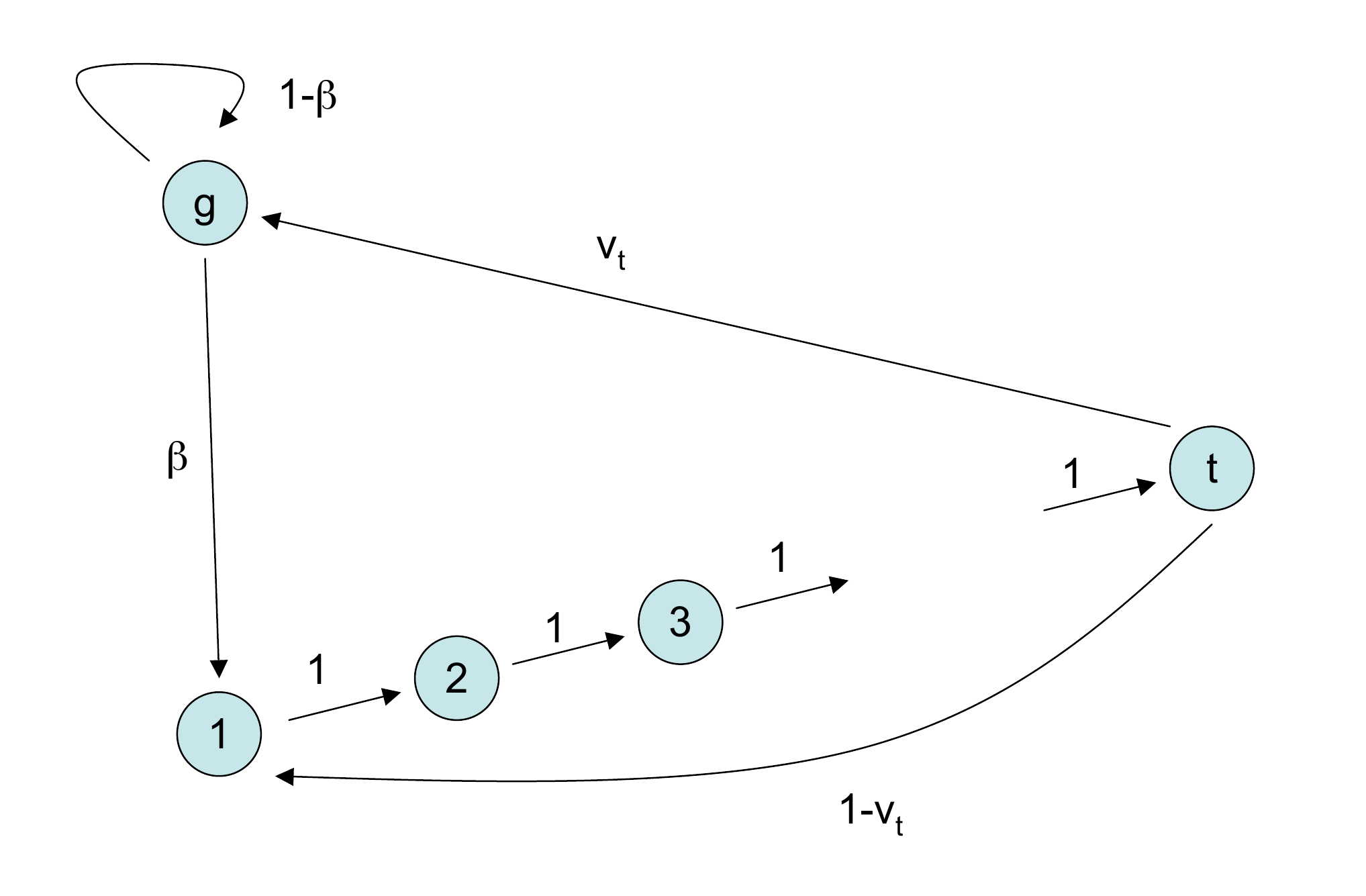}
}
\caption{\label{fig:pkpic} Markov Chain for policy $\P(t)$.}
\end{figure}

\begin{lemma}
\label{lem:main1} 
In playing an arm with reward $r$, transition probabilities $\alpha$ and $\beta$, the policy $\P(t)$ yields average reward $R(t) = r \frac{v_t}{v_t+t \beta}$, and expected rate of play $Q(t) = \frac{v_t + \beta}{v_t + t \beta} \ge \frac{1}{t}$. Recall that $v_t=\frac{\alpha}{\alpha+\beta}(1-(1-\alpha-\beta)^t)$ is the probability the arm is good given it was observed to be bad $t$ steps ago.
\end{lemma}
\begin{proof}
  The Markov chain describing the policy $\P(t)$  is shown in
  Figure~\ref{fig:pkpic}, and has $t+1$ states
  which we denote $s,0,1,2,\ldots,t-1$. The state $s$ corresponds to
  the arm being observed to be in state $g$, and the state $j$
  corresponds to the arm being observed in state $b$ exactly $j$ steps
  ago. The transition probability from state $j$ to state $j+1$ is
  $1$, from state $s$ to state $0$ is $\beta$, from state $t-1$ to
  state $s$ is $v_{t}$, and from state $t$ to state $0$ is
  $1-v_{t}$. Let $\pi_s, \pi_0, \pi_1, \ldots, \pi_{t-1}$ denote the
  steady state probabilities of being in states $s,0,1,\ldots,t-1$
  respectively.  This Markov chain is easy to solve. We have $\pi_0 =
  \pi_1 \ldots = \pi_{t-1}$, so that the first identity is: $\pi_s + t
  \pi_0 = 1$.  Furthermore, by considering transitions into and out of
  $s$, we obtain: $ \beta \pi_s = v_t \pi_{t-1} = v_t \pi_0$.
  Combining these, we obtain: $ \pi_s = \frac{v_t}{v_t + t \beta}$,
  and $\pi_0 = \frac{\beta}{v_t + t \beta} $.  Now we have:
 $$R(t) = r[(1-\beta) \pi_s + v_t \pi_0] = r\pi_s = r\frac{v_t}{v_t + t \beta}$$
 $$ Q(t) = \pi_s + \pi_{t-1} = \frac{v_t + \beta}{v_t + t \beta}$$
 \end{proof}

\begin{lemma}
\label{lem:poly} (Proved in Appendix~\ref{app:omitted}) For each arm $i$, the optimal reward minus penalty of the single arm policy for arm $i$ is 
$$ H_i(\lambda) = \max_{t \ge 1} F_i(\lambda,t) = \max_{t \ge 1} \left( \frac{(r_i-\lambda) v_{it} - \lambda \beta_i}{v_{it} + t \beta_i} \right)$$
The maximum value $t_i(\lambda) = \mbox{argmax}_{t \ge 1} F_i(\lambda,t)$ satisfies the following:
\begin{enumerate}
\item If $\lambda \ge r_i \left(\frac{\alpha_i}{\alpha_i + \beta_i(\alpha_i+\beta_i)} \right)$, then $t_i(\lambda) =\infty$, and $H_i(\lambda) = 0$.
\item If $\lambda = r_i \left(\frac{\alpha_i}{\alpha_i + \beta_i(\alpha_i+\beta_i)} \right) -\rho $ for some $\rho > 0$, then
$t_i(\lambda)$ (and hence $H_i(\lambda)$) can be computed in time polynomial in the input size
and in $\log(1/\rho)$ by binary search.
\end{enumerate}
\end{lemma}

\newcommand{\EE}{\epsilon}

\subsection{The {\sc BalancedIndex} Policy}
\label{sec:bala}
Though we could now use {\sc LPLagrange}$(\lambda)$ to solve Whittle's LP by finding the $\lambda$ so that $\sum_{i=1}^n Q_i(\lambda) \approx 1$ (refer Appendix~\ref{app:lp} for details), our $2$-approximation policy will {\em not} be based this approach. For our analysis to work, we must make a subtle but crucial modification: We will instead set $\lambda$ to be the sum of the excess reward for all single-arm policies $\sum_{i=1}^n H_i(\lambda)$. (Recall that we can interpret $\lambda$ to be a penalty per play, so in the optimal single-arm policy for arm $i$, $H_i(\lambda)$ is the average reward minus penalty.) Note that by Lemma~\ref{lem:lagcrux}, this implies $\lambda \ge OPT/2$ and $\sum_{i=1}^n H_i(\lambda) \ge OPT/2$. Intuitively, we are forcing the Lagrangean to {\em balance} short-term reward (represented by $\lambda$) with long-term average reward (represented by $\sum_{i=1}^n H_i(\lambda)$). Our {\em balance} technique can be generalizes to many other restless bandit problems (see Sections~\ref{sec:recover} -- \ref{sec:beyond}).

We first show how to compute this value of $\lambda$ in polynomial time. We begin by presenting the connection between $G(\lambda = \sum_{i=1}^n H_i(\lambda)$ and $OPT$, the value of the optimal solution to {\sc (Whittle)}.

\begin{lemma}
\label{lem:lagcrux}
For any $\lambda$, we have: $\lambda + G(\lambda)  = \lambda + \sum_{i=1}^n H_i(\lambda)  \ge OPT$.
\end{lemma}
\begin{proof}
 By Lemma~\ref{lem:approxbalance}, part (1), we have: $ \lambda + \sum_{i=1}^n H_i(\lambda)  = \lambda + \sum_{i=1}^n h_i$. The latter is the objective of the dual of {\sc (Whittle)}, which implies the lemma by weak duality.
\end{proof}

\begin{lemma}
\label{lem:mon}
$h_i = H_i(\lambda)$ is a non-increasing function of $\lambda$.
\end{lemma}
\begin{proof}
 Recall from Lemma~\ref{lem:approxbalance} that $h_i = H_i(\lambda)$. For any $\lambda$, consider the value $F_i(\lambda,t)=R_i(t)-\lambda Q_i(t)$ of the policy $\P_i(t)$. Since this decreases as $\lambda$ increases, $H_i(\lambda) = \max_t F_i(\lambda,t)$ is also a non-increasing function of $\lambda$.
\end{proof}

\begin{lemma}
\label{lem:poly2}
In polynomial time, we can find a $\lambda$ so that $ \lambda \ge (1-\EE) OPT/2$, and $G(\lambda) = \sum_{i=1}^n H_i(\lambda) = \sum_{i=1}^n h_i  \ge OPT/2$.
\end{lemma}
\begin{proof}
First note by Lemma~\ref{lem:mon} that  $G(\lambda) = \sum_{i=1}^n H_i(\lambda)$ is monotonically non-increasing in $\lambda$. Therefore, start with $\lambda = \sum_{i=1}^n r_i$, and $\lambda$ scale down by by a factor of $(1+\EE)$ until $\lambda < G(\lambda)$. Note that for any $\lambda$, the value of $G(\lambda)$ can be computed in poly-time by Lemma~\ref{lem:poly}.  At this point, let  $\lambda' = \lambda(1+\EE)$. Since $G(\lambda') \le \lambda'$, by Lemma~\ref{lem:lagcrux}, we have $\lambda' \ge OPT/2$, which implies $\lambda \ge (1-\EE) OPT/2$. Further, since $\lambda < G(\lambda)$, again by Lemma~\ref{lem:lagcrux}, we have $G(\lambda) \ge OPT/2$.
\end{proof}

\subsubsection{Index Policy}
We start with the value of $\lambda$ from Lemma~\ref{lem:poly2}. The policy only works with the subset of arms $S$ so that for $i \in S$, we have $H_i(\lambda) > 0$. For this $\lambda$, the solution to {\sc LPLagrange}$(\lambda)$ yields one policy $\P_i(t_i(\lambda))$  of value $H_i(\lambda)$ for each arm $i \in S$ (see Lemma~\ref{lem:approxbalance}). Let $t_i = t_i(\lambda)$. Recall that if an arm was last observed in state $s \in \{g,b\}$ some $t \ge 1$ steps ago, then its state is denoted $(s,t)$.  We call an arm $i$ in state $(g,1)$ as {\em good}; in state $(b,t)$ for $t \ge t_i$ as {\em ready}, and in state $(b,t)$ for $ t < t_i$ as {\em bad}.   The policy is  shown in Figure~\ref{fig:combine}. 

\medskip
\begin{figure}[htbp]
\centerline{\framebox{\small
\begin{minipage}{4.5in}
{\bf {\sc BalancedIndex} Policy for {\sc Feedback} MAB}
\begin{tabbing} 
Consider only arms with $H_i(\lambda) > 0$; denote these as set $S$.\\
Play arms in $S$ according to the following priority scheme: \\
1. \= {\bf Exploit:} Play any arm in state $(g,1)$ ({\em good} state): \\ 
2. \> If condition (1) does not hold: \\
\>  \ \ \= (a) \ \= {\bf Explore:} Play any arm $i \in S$ in state $(b,t)$ such that $t \ge t_i$  ({\em ready} state). \\ 
\>  \> (b) \>   {\bf Idle:} If no arm is {\em good} or {\em ready} (all arms {\em bad}), do not play at this step.
\end{tabbing}
\end{minipage}
}}
\caption{\label{fig:combine}\label{fig:feedindex} The {\sc BalancedIndex} Policy for {\sc Feedback} MAB.}
\end{figure}

Note that the way the scheme works, at most one arm can be in state $(g,1)$ at any time step, and if such an arm exists, this arm is played at the current step (and in the future until it switches out of this state). The above can be thought of as executing the policies $\P_i(t_i)$ for arms $i \in S$ independently and in case of simultaneous attempts to play, resolving conflicts according to the above priority scheme.

Though the above policy is not written as an index policy, it is equivalent to the following index: There is a dummy arm with index $0$ that does not yield reward on playing.  If $h_i = H_i(\lambda)= 0$, the index for all states of this arm is $-1$. For arms with $h_i > 0$, the index for state $(g,1)$ is $2$; that for states $(b,t)$ with $t \ge t_i$ is $1$, and that for states $(b,t)$ with $t < t_i$ is $-1$.

\subsubsection{Analysis}
We now prove that the {\sc BalancedIndex} policy is in fact a 2-approximation. The proof is based on the fact that the Lagragean $\lambda$ and the excess rewards $h_i=H_i(\lambda)$ give us a way of accounting for the average reward. And by Lemma~\ref{lem:lagcrux}, $\lambda \ge OPT/2$ and $\sum h_i \ge OPT/2$, which gives us a way of linking the rewards from our policy to the LP optimum.

\begin{theorem}
\label{thm:bal}
The {\sc BalancedIndex} policy is a $2+\EE$ approximation to {\sc Feedback MAB}. Furthermore, this policy can be computed in polynomial time.
\end{theorem}

\begin{proof}
Recall that the reward of optimal single arm policy $\P_i(t_i)$ is $R_i(\lambda) = H_i(\lambda) + \lambda Q_i(\lambda)$, so that this reward can be accounted as $H_i(\lambda)=h_i$ per step plus $\lambda$ per play. We use this amortization of rewards to show that the average reward of our index policy is at least $OPT/2$.

Focus on any arm $i$, we call a step {\em blocked} for the arm if the arm is ready for play--the state is $(b,t)$ where $t \ge t_i$--but some other arm is played at the current step. Consider only the time steps which are {\em not} blocked for arm $i$. For these time steps, the arm behaves as follows: It is continuously played in state $(g,1)$. Then it transitions to state $(b,1)$ and moves in $t_i-1$ time steps to state $(b,t_i-1)$. After this the arm might be blocked, and the next state that is not blocked is $(b,t)$ for some $t \ge t_i$, at which point the arm is played. Using the formula for $R(t)$ from Lemma~\ref{lem:main1}, and since $v_{it} \ge v_{it_i}$ for $t \ge t_i$, we have
 
$$ R_i(t) \ge r_i \frac{v_{it}}{v_{it} + t_i \beta_i} \ge r_i \frac{v_{it_i}}{v_{it_i} + t_i \beta_i} = R_i(t_i)$$

which implies that the per-step reward of this single arm policy for arm $i$ restricted to the non-blocked time steps is at least the per-step reward $R_i(t_i)$ of the optimal single-arm policy $\P_i(t_i)$. Therefore, for these non-blocked steps, the reward we get is at least $h_i = H_i(\lambda)$ per step, and at least $\lambda$ per play. 

Now, on steps where no arm is played, none of the arms is blocked by definition, so our amortization yields a per-step reward of at least $\sum_{i \in S} h_i \ge OPT/2$. On steps when some arm is played, the arm that is played by definition cannot not blocked, so we get a reward of at least $\lambda \ge (1-\EE) OPT/2$ for this step. This completes the proof. 
\end{proof}

\subsubsection{Alternate Analysis}
\label{sec:analysis}
The above analysis is very intuitive. We now present an alternative way to analyze the policy, that leads to a more generalizable technique. This uses a Lyapunov (potential) function argument. Recall from Lemma~\ref{lem:approxbalance} that $h_i = H_i(\lambda)$; further that $t_i = t_i(\lambda)$. Define the potential $\Phi_i$ for each arm $i$ at any time as follows: 
 
\begin{define}
If arm $i$ moved to state $b$ some $y$ steps ago ($y \ge 1$), the potential $\Phi_i$ is $ h_i (\min(y,t_i)-1) $. In the
state $g_i$ the potential is $p_i$. Recall that $p_i$ is the optimal dual variable in {\sc Whittle-Dual}$(\lambda)$.
\end{define} 

Let $\Phi_T$ denote the total potential, $\sum_{i=1}^n \Phi_i$, at any step $T$ and let $R_T$ denote the total reward accrued until that step. Define the function $\L_T = t \cdot OPT/2 - R_T - \Phi_T$. Let $\Delta R_T = R_{T+1} - R_T$ and $\Delta \Phi_T =
\Phi_{T+1} - \Phi_T$. 

\begin{lemma}
\label{lem:crux1}
 $\L_T$ is a Lyapunov function. {\em i.e.}, $\E[\L_{T+1} |\L_T] \le \L_T$. Equivalently, at any step:
$$ \E[\Delta R_T + \Delta \Phi_T|R_T,\Phi_T] \ge  (1-\EE) OPT/2$$
\end{lemma}
\begin{proof} 
At a given step, suppose the policy does nothing, then all arms are ``not
ready". The total increase in potential is precisely 
$$\Delta \Phi_T = \sum_{i \in S} h_i = G(\lambda) \ge OPT/2$$

On the other hand, suppose that the policy plays arm $i$, which has last
been observed in state $b$ and has been in that state for $y \ge t_i$ steps.
With probability $q \geq v_{it_i}$ the observed state is $g_i$ and 
the change in reward $\Delta R_T=r_i$ and the change in potential is $p_i
- h_i (t_i-1)$. With probability $1-q$ the observed state is $b$ and the
change in potential is $ - h_i (t_i-1)$ (and there is no change in reward).
Thus in this case since $q \geq v_{it_i}$, and $p_i \ge 0$, we have:
\[ \E[\Delta R_T + \Delta \Phi_T|R_T,\Phi_T] \ge q p_i - h_i (t_i-1)
\geq v_{it_i} p_i - h_i(t_i -1) \geq  \lambda + h_i \ge (1-\EE) OPT/2 \]
The penultimate inequality follows from Lemma~\ref{lem:approxbalance}, part (3). Note that the potentials of arms not played cannot decrease, so that the first inequality is valid.

Finally supposing the policy plays an arm $i$ which was last observed in
state $g_i$ and played in the last step, with probability $1-\beta_i$ the
increase in reward is $r_i$ and the potential is unchanged. With
probability $\beta_i$ the potential will decrease by $ -p_i$. Therefore in
this case, by Lemma~\ref{lem:approxbalance}, part (4). 
\[ \E[\Delta R_T + \Delta \Phi_T|R_T,\Phi_T] \ge r_i - \beta_i 
p_i \geq  \lambda + h_i \ge (1-\EE) OPT/2 \] 
\end{proof}

By their definition, the potentials $\Phi_T$ are bounded independent of
the time horizon, by telescoping summation, the above lemma implies that
$\lim_{T \rightarrow \infty} \frac{\E[R_T]}{T} \ge (1-\epsilon)OPT/2$.  This proves Theorem~\ref{thm:bal}.

\paragraph{Gap of Whittle's LP.} The following theorem shows that our analysis is almost tight (considering that our $2$-approximation is against Whittle's LP).

\begin{theorem}
\label{thm:whittle-gap}
(Proved in Appendix~\ref{app:omitted}) The  gap of Whittle's LP is arbitrarily  close to $e/(e-1)\approx 1.58$.
\end{theorem}

\section{Analyzing the Whittle Index for {\sc Feedback} MAB}
Before generalizing our $2$-approximation algorithm to a larger subclass of restless bandit problems, we explore the connection between our analysis and the well-known Whittle Index used in practice. This section can be skipped without losing continuity of the paper.

\label{sec:index}
A well-studied index policy for restless bandit problems is the {\em
  Whittle Index}~\cite{whittle2}. In the context of {\sc Feedback MAB},
this index has been independently studied by Le Ny {\em et al}~\cite{leny}
and subsequently by Liu and Zhao~\cite{zhao2}. Both these works give a
closed form expressions for this index and show near-optimal empirical
performance. Our main result in this section is to justify the empirical
performance by showing that a simple but very natural modification of this
index in order to favor myopic exploitation yields a $2$-approximation.
The modification simply involves giving additional priority to arms in
state $(g,1)$ if their myopic expected next step reward $r_i (1-\beta_i)$
is at least a threshold value.

\subsection{Description of the Whittle Index} 
\label{sec:whitindex}
Defined in general, the Whittle's index for each state $x$ is the largest penalty-per-play $\lambda$ such that the optimal policy is indifferent between playing in $x$ and not playing. In our specific problem, the current state for each arm $i$ is captured by the tuple $(s,t)$-- the arm was last seen to be $s \in \{g,b\}$ (good or bad) $t$ steps ago. The Whittle index $\Pi_i(s,t)$ is a non-negative real numbers computed as follows: using the notation from Section~\ref{app:gap2}., for any penalty per play $\lambda$, there is a single-arm policy $L_i(\lambda)$  that maximizes the average reward minus penalty (excess reward) $H_i(\lambda)$ over the infinite horizon. When $\lambda=\infty$, the optimal policy never plays; when $\lambda=0$, the optimal policy would play in any state. As $\lambda$ is decreased from $\infty$, at some value $\lambda^*$, the decision in state $(s,t)$ changes from ``not play" to ``play". The Whittle index $\Pi_i(s,t)$ is precisely this value of  $\lambda^*$.  The Whittle index policy always plays the arm with the highest Whittle's index (Fig.~\ref{fig:ind-wh}).

\begin{figure}[htbp]
\centerline{\framebox{\small
\begin{minipage}{5.2in}
{\bf Whittle Index Policy:} Play the arm  $i$ whose current state $(s,t)$ has the highest index $\Pi_i(s,t)$.
\end{minipage}
}}
\caption{\label{fig:ind-wh} The Whittle Index Policy.}
\end{figure}

\paragraph{Remarks.} The Whittle index is strongly decomposable, {\em i.e.}, can be computed separately for each arm.  Further, we have defined $\lambda$ as a penalty (or amortized reward) per play, while Whittle defines it as a reward for not playing (which he terms the {\em subsidy for passivity}); it is easy to see that both these formulations are equivalent. Finally, for {\sc Feedback MAB}, it can be shown~\cite{leny,zhao2} that for any state $(s,t)$, there is  a {\em unique} $\lambda \in (-\infty,\infty)$ where the decision switches between ``play" and ``not play", {\em i.e.}, the decision is {\em monotone} in $\lambda$. Strictly speaking, the Whittle index is defined only for such systems (termed {\em indexable} by Whittle~\cite{whittle2}); we will define this aspect away by insisting that the index $\lambda^*$ is the {\em largest}  value where a switch happens.

\medskip

We present an explicit  connection of Whittle's index to {\sc LPLagrange}$(\lambda)$. 
\begin{lemma}
\label{lem:whit}
 (Proved in Appendix~\ref{app:omitted}) Recall the notation $L_i(\lambda)$ and $\P_i(t)$ from Section~\ref{app:gap2}. The following hold for $\Pi_i(s,t)$:
\begin{enumerate}
\item $\Pi_i(s,t) \ge 0$ for all states $(s,t)$ where $s \in \{g,b\}$ and $t \in \mathbf{Z}^+$.
\item $\Pi_i(g, 1) = r_i (1-\beta_i)$, and $\Pi_i(b,t) \le \Pi_i(g,1)$ for all $t \ge 1$.
\item $\Pi_i(b,t) = \max\{ \lambda | L_i(\lambda) = \P_i(t)\}$, and is a monotonically non-decreasing function of $t$.
\end{enumerate}
\end{lemma}

Though Whittle's index is widely used, it is not clear how to analyze it since it leads to complicated priorities between arms. We now show that our balancing  technique also implies an analysis for a slight but non-trivial modification to Whittle's index.

\subsection{The {\sc Threshold-Whittle} Policy}
\label{sec:threshindex}
We now show that modifying the index  slightly to exploit the myopic next step reward in good states $g$ yields a $2$ approximation. Note that the myopic next step reward of an arm $i$ in state $g$ is precisely $\Pi_i(g,1) = r_i(1-\beta_i)$. The modification essentially favors exploiting such a ``good'' state if the myopic reward is at least a certain threshold value.  In particular, we analyze the policy {\sc Threshold-Whittle}$(\lambda)$ shown in Figure~\ref{fig:whitindex}, where we set $\lambda = \lambda^*$, where $\lambda^*$ is the value where $\lambda^* = \sum_{i=1}^n H_i(\lambda^*)$ (refer Section~\ref{sec:bala}). 

\medskip
\begin{figure}[htbp]
\centerline{\framebox{\small
\begin{minipage}{5.0in}
{\sc Threshold-Whittle}$(\lambda)$
\begin{tabbing}
{\bf At} any time step: \\
\ \ \ \= {\bf If} $\exists$ arm $i$ in state $(g,1)$ whose Whittle index is $\Pi_i(g,1) = r_i (1-\beta_i) \ge \lambda$ \\
\> \ \ \ \ \={\bf then} Play arm $i$. \\
\> \> {\bf else} Play the arm with the highest Whittle index.
\end{tabbing}
\end{minipage}
}}
\caption{\label{fig:whitindex} Policy {\sc Threshold-Whittle}$(\lambda)$. It exploits arm $i$ if the myopic reward in state $(g,1)$ is $ \ge \lambda$.}
\end{figure}

Note that the above policy can be restated as playing the arm with the highest modified index, which is computed as follows: For arm  $i$, if $\Pi_i(g,1) = r_i(1-\beta_i) \ge \lambda$, the modified index for  state $(g,1)$ is  $\infty$, else the modified index is the same as the Whittle index.

\begin{theorem}
  \label{thm:whittle}
 {\sc Threshold-Whittle}$(\lambda^*)$ is a $2$ approximation for {\sc Feedback MAB}. Here, $\lambda^*$ satisfies $\lambda^* = \sum_{i=1}^n H_i(\lambda^*)$ (refer Section~\ref{sec:bala}).
 \end{theorem}

\subsection{Proof of Theorem~\ref{thm:whittle}}
\label{sec:connections}
We now prove the above result by modifying our analysis of the {\sc BalancedIndex} policy (from 
Figure~\ref{fig:feedindex}). Recall that $S$ is the set of arms with
$h_i > 0$ in the optimal solution to {\sc Whittle-Dual}$(\lambda^*)$. For such arms, $t =
t_i$ is the first time instant when $\lambda + th_i \ge p_i v_{it}$ is
tight. For arm $i \in S$, state $(s,t)$ is {\em good} if $s = g$ and
$t=1$; {\em ready} if $s = b$ and $t \ge t_i$; and {\em bad} otherwise.
The index policy from Figure~\ref{fig:feedindex} favors {\em good} over
{\em ready} states, and does not play any arm in {\em bad} states.

\begin{claim}
  \label{claim:whit}
  For any arm $i$, exactly one of the following is true for {\sc Whittle-Dual}$(\lambda^*)$ and {\sc LPLagrange}$(\lambda^*)$.
  \begin{enumerate}
  \item The constraint $\lambda^* + t h_i  \ge   v_{it} p_i $ is first tight at $t = t_i$.   Then, $\Pi_i(b,t_i-1) < \lambda^*$ and $\Pi_i(b,t_i) \ge \lambda^*$.    Further, $\Pi_i(g,1) = r_i (1-\beta_i) \ge \lambda^*$ and $h_i > 0$.
  \item The constraint $\lambda^* + t h_i  \ge   v_{it} p_i $ is not tight for any $t$. Then,    $\Pi_i(b,t) \le \lambda^*$ for all $t \ge 1$,  and $h_i = 0$.
  \end{enumerate}
\end{claim}
\begin{proof}
The optimal solution to {\sc LPLagrange}$(\lambda^*)$ finds the policy $\P_i(t_i)$ for every arm $i$ with $h_i > 0$. Therefore, by Lemma~\ref{lem:whit}, we must have $\Pi_i(b,t_i) \ge \lambda^*$, and  $\Pi_i(b,t) < \lambda^*$ for all $t < t_i$. Furthermore, since the variable $x^i_{bt}$ in the optimal solution to  {\sc LPLagrange}$(\lambda^*)$ is first non-zero at $t = t_i$, this implies the constraint $\lambda^* + t h_i  \ge   v_{it} p_i $ is first tight at $t = t_i$ by complementary slackness (Lemma~\ref{lem:approxbalance}).  Further, if this constraint is tight at $t = t_i$, since $v_{it}$ is  monotonically increasing, the constraint is feasible for all $t \ge t_i$ only if $h_i > 0$. Finally,  $\Pi_i(g,1) = r_i (1-\beta_i) \ge \Pi_i(b,t_i) \ge  \lambda^*$ follows from Lemma~\ref{lem:whit}. 

Suppose now that  $\lambda^* + t h_i  \ge   v_{it} p_i $ is not tight for any $t \ge 1$. Then, by complementary slackness, we have $x^i_{bt} = 0$ for all $t \ge 1$, which implies $x^i_{gt} = 0$ for all $t \ge 1$. Therefore, the policy $L_i(\lambda^*)$  never plays arm $i$. This implies  $\Pi_i(b,t) \le \lambda^*$ for all $t \ge 1$. Since  the excess reward of $L_i(\lambda^*)$ is zero, we have $h_i = 0$. (This can also be shown by complementary slackness.)
\end{proof}

\subsubsection{Types of Arms}
We next classify the arms as follows. In Claim~\ref{claim:whit}, let the arms satisfying the first condition ($h_i > 0$) of the Claim be denoted Type (1), and the remaining arms satisfying $h_i = 0$ be denoted Type (2). Note that type (1) arms are precisely the set $S$ of arms in Fig.~\ref{fig:feedindex}, so the {\sc BalancedIndex} policy only plays type (1) arms.

\paragraph{Type (1): Arms in $S$.}  We  consider the behavior of {\sc Threshold-Whittle}$(\lambda^*)$ restricted to just these arms. Since $\Pi_i(b, t)$ is monotonically increasing in $t$, by Claim~\ref{claim:whit}, we have the following for  the policy of Fig.~\ref{fig:feedindex}:  If the arm is {\em ready}, the Whittle index is at least $\lambda^*$;  if the arm is {\em bad}, the index is at most $\lambda^*$; and finally, if the arm is {\em good}, then the modified Whittle index is infinity.

Therefore, {\sc Threshold-Whittle}$(\lambda^*)$ confined to these arms gives priority to {\em good} over {\em ready} over {\em bad} arms. The only difference with the policy in Fig.~\ref{fig:feedindex} is that instead of idling when all arms are {\em bad}, the policy  {\sc Threshold-Whittle}$(\lambda^*)$ will play some {\em bad} arm.  We now show that this is  better than idling.

\begin{claim}
  \label{claim:priority}
 {\sc Threshold-Whittle}$(\lambda^*)$ executed just over Type (1) arms yields   reward at least $OPT/2$.
\end{claim}
\begin{proof}
Consider the alternate analysis presented in Section~\ref{sec:analysis}.  The {\sc Index} policy from Fig.~\ref{fig:feedindex} does not play an   arm $i$ in {\em bad} state, and achieves change in potential $\Delta \Phi$ of exactly $h_i$. All we need to show is that if the arm is  played instead, the expected change in potential is still at least $h_i$. The rest   of the proof is the same as that of Lemma~\ref{lem:crux1}.  Suppose  the arm is played after $t \ge 1$ steps. The expected change in  potential is: $ \E[\Delta \Phi_t] = v_{it} p_i - h_i(t -1)$.  We  further have by definition of $t_i$ that $\lambda + t_i h_i  = v_{it_i} p_i$. We therefore have $p_i v_{it_i} \ge t_i h_i$. Since  $v_{it}$ is a concave function of $t$ with $v_{i0} = 0$, the above  implies that for every $t \le t_i$, we must have $p_i v_{it} \ge t  h_i$.  Therefore, $ \E[\Delta \Phi_t] = v_{it} p_i - h_i(t -1) \ge t  h_i - h_i(t -1) = h_i$.
\end{proof}

\paragraph{Type (2): Arms not in $S$.} The only catch now is that  {\sc Threshold-Whittle}$(\lambda^*)$ can sometimes play a type (2)  arm whose $h_i = 0$. For such arms, we count their reward and ignore the change in potential.

\begin{lemma}
In  {\sc Threshold-Whittle}$(\lambda^*)$, if a type (2) arm $j$ preempts the play of a type (1) arm $i$, either the reward from the former is at least $\lambda^*$ or the increase in potential of the later $\Delta \Phi$ is at least $h_i$.
\end{lemma}
\begin{proof}
Suppose that for type (2) arm $j$, $\Pi_j(g,1) = r_j(1-\beta_j) \ge \lambda^*$. Denote such a state $(g,1)$ as {\em nice}, and a {\em nice} type (2) arm has modified index of $\infty$. When $j$ was last observed to be good, the arm can be played continuously even if type (1) arms become {\em ready}. However, for every time step such an event happens, the current reward of playing this {\em nice} type (2) arm is precisely $r_j(1-\beta_j)$, which is at least $\lambda^*$, and the type (1) arms only get better from waiting. When the type (2) arm $j$ was last observed to be bad, preemption can only happen if all type (1) arms are bad, since the Whittle's index of a type (2) arm $\Pi_j(b,\infty) < \lambda^*$. But in this case the increase in potential of arm $i$ is $h_i$. 

Finally, if $\Pi_j(g,1) < \lambda^*$, then by Lemma~\ref{lem:whit}, we have $\Pi_j(s,t) < \lambda^*$ for all $s = b, g$ and $t \ge 1$. This implies that such an arm in any state can only preempt type (1) arms that are {\em bad}; in that case, the  potential $\Phi$ of the latter rises by $h_i$ by idling. This completes the proof.
\end{proof}

\subsubsection{Completing the Proof of Theorem~\ref{thm:whittle}}

To complete the analysis, there are two cases: First, if a {\em nice} type
(2) arm, or a {\em ready} or {\em good} type (1) arm is played, then the
above discussion implies that the reward plus change in potential ($\Delta
\Phi$) of this arm is at least $\lambda^* \ge OPT/2$.  In the
other case, all type (1) arms are {\em bad}, and focusing on just these
arms, each yields increase in potential for each arm is at least $h_i$, so
that the total reward plus change in potential of these system is at least
$\sum_i h_i \ge OPT/2$. This completes the proof, and shows that {\sc
  Threshold-Whittle}$(\lambda^*)$ is a $2$ approximation.
We note that the above analysis extends easily to the variant where $M \ge
1$ arms are simultaneously played per step.

\section{The General Technique: {\sc Monotone} Bandits}
\label{sec:recover}
In this section, we present a general and non-trivial sub-class of restless bandits for which a generalization of the above balancing technique yields a $2$-approximate index policy.  We term this class {\sc Monotone} bandits, and this captures both the stochastic MAB, as well as the {\sc Feedback} MAB as special cases. 

In {\sc Monotone} bandits, there are $n$ bandit arms.  Each arm $i$ can be in one of $K$ states denoted $\S_i = \{\sigma^i_1, \sigma^i_2, \ldots, \sigma^i_K\}$.   
When the arm is not played, its state remains the same and it does not
fetch reward.  Suppose the arm is in state $\sigma_k^i$ and is played next
after $t \ge 1$ steps. Then, it gains reward $r_k^i \ge 0$, and
transitions to one of the states $\sigma_j^i \neq \sigma_k^i$ w.p.
$g^i(k,j,t)$, and with the remaining probability stays in state
$\sigma^i_k$.  (For notational convenience, we denote $\sigma^i_k$ simply
as $k$; the arm it refers to will be clear from the context.) The transition probabilities for different arms are independent. At most one arm is played per step.  The goal is to find a policy for
playing the arms so that the infinite horizon time-average reward is
maximized.

In addition, we have the following key properties about the transition probabilities:

\begin{description}
\item[Separability Property:] We assume that $g^i(k,j,t)$ is of the
form $f^i_k(t) q^i(k,j)$. The function $f^i_k(t) \in [0,1]$ for positive integers $t$
can be thought of as an ``escape probability'' from the state $\sigma_k
\in \S_i$. Conditioned of the escape, the state changes to $\sigma_j \in \S_i$ with
probability $q^i(k,j)$, thus $\sum_{j \ne k} q^i(k,j) \le 1$.

\item[Monotone Property:]  For every arm $i$ and state $k \in
\S_i$, we have: $f_k^i(t) \le f_k^i(t+1)$ for every $t$. 
\end{description}

The above properties are necessary in some sense: We show in Section~\ref{app:hard} that when the monotone property
is relaxed, the problem becomes $n^{\epsilon}$-hard to approximate.
Further, if the separability property is not satisfied, then Whittle's LP
on which the analysis of this section is based, has $\Omega(n)$ gap.

\paragraph{Motivation and Special Cases.} Intuitively, {\sc Monotone} bandit models optimization scenarios in which uncertainty increases: when an arm is just played and we observe its state, we are most certain that our observation still holds true the next time step. However, the non-decreasing nature of $f$ implies that as time goes on, the ``escape probability" increases and the previous observation becomes less and less reliable.  This serves as a model for certain  POMDPs, such as the {\sc Feedback} MAB.

Observe that the {\sc Monotone} bandits generalizes the 
{\sc Feedback} MAB. For the states $\S_i = \{g,b\}$, set $q^i(g,b)=q^i(b,g)=1$ and
$f^i_g(t) = 1 - u_{it}$ and $f^i_b(t)=v_{it}$. Recall from Fact~\ref{fact1} that $u_{it},v_{it}$
are respectively the probabilities of observing the state $g$ when the state last
observed $t$ steps ago was $g$ and $b$, and that
$1-u_{it},v_{it}$ are both monotonically increasing.  We also note that {\sc Monotone} bandits generalizes  the stochastic MAB  by setting $f^i_k(t)=1$ for all $t$.

\subsection{High Level Idea}
Unlike the {\sc Feedback} MAB problem, in {\sc Monotone} bandits, there is no longer a clear distinction between ``good" and ``bad" states. Note however that an equivalent way of finding $\lambda$ such that $\lambda = \sum_{i=1}^n H_i(\lambda)$ is to treat $\lambda$ as a variable and enforce $\lambda = \sum_{i=1}^n h_i$ as a constraint in the dual of Whittle's LP. By taking this approach, the variables $p^i_k$ (now one for each state $k \in \S_i$) can be interpreted as dual potentials, and the dual constraints are in terms of the expected potential change of playing in state $k \in \S_i$. Based on the sign of this potential change, we can classify the states into ``good" and ``bad" via complementary slackness. Our index policy continuously exploits arms in ``good" states, and waits until the dual constraint goes tight ({\em i.e.}, the arm becomes ``ready") before playing in ``bad" states. We formalize the previous potential-based argument using a Lyapunov function and show a $2$-approximation. We note that the LP-duality approach is entirely equivalent to the Lagrangean approach; however, it leads to a different interpretation of variables which is more generalizable.

\paragraph{Technical Assumptions.} For simplicity of the exposition, we assume the monotone functions in this
section are piece-wise linear with poly-size specification -- see
definition~\ref{def:break} for a formal definition. As shown in the previous section, these results do extend
to a wider class of differentiable functions, such as those in
{\sc Feedback} MAB.

We also assume that for each arm $i$, the graph, where the vertices
are $k \in \S_i$ and a directed edge $(j,k)$ exists if $q^i(j,k) > 0$, is
strongly connected. Since we consider the infinite horizon time average
reward, assume that the policy is ergodic and can choose the start state
of each arm.  These assumptions do not simplify the problem, as it remains {\sc NP-Hard} (see 
Section~\ref{app:hard}).

\subsection{Whittle's LP and its Dual}
\label{sec:whit}
\eat{We first present the linear programming relaxation due to
Whittle~\cite{whittle2}, and take its dual.  We {\em do not} solve this
relaxation. Instead, we solve the dual of a slightly different relaxation
which we present in Section~\ref{sec:balance}. 
This section introduces a lot of notation; we summarize the notation in Table~\ref{tab1}, and present an interpretation along with a roadmap of the analysis in Section~\ref{sec:intuition}. }

As with {\sc Feedback MAB}, for each arm $i$ and $k \in \S_i$, we have variables $\{x^i_{kt}, t \ge 1\}$.  These variables capture the probabilities (in the execution of the optimal policy) of the event:
 Arm $i$ is in state $k$, was last played $t$ steps ago, and is
played at the current step. These quantities are well-defined for
ergodic policies. Whittle's LP is presented in Figure~\ref{fig:monobasic}. Let its optimal value
be denoted $OPT$.  The LP effectively encodes the constraints on the
evolution of the state of each arm separately, connecting them only by the
constraint that at most one arm is played {\em in expectation} every step. The first constraint simply states that the at any step, at most one arm is played; the second constraint encodes that each arm can be in at most one possible state at any time step; and the final constraint encodes that the rate of entering state $k \in \S_i$ is the same as the rate of exiting this state. This LP will clearly be a
relaxation of the optimal policy; the details are the same as the proof of Lemma~\ref{lem:optopt}.

\begin{figure*}[htbp]
\centerline{\framebox{\small
\begin{minipage}{6in}
\[ \mbox{Maximize       } \ \ \ \  \sum_{i=1}^n\sum_{k \in \S_i} \sum_{t \ge 1} r^i_k x^i_{kt} \qquad \qquad \mbox{\sc (Whittle)}  \] 
\[ \begin{array}{rcll}
\sum_{i=1}^n \sum_{k \in \S_i} \sum_{t \ge 1} x^i_{kt} & \le & 1 & \\
\sum_{k \in \S_i} \sum_{t \ge 1} t x^i_{kt} & \le & 1 & \forall i \\
\sum_{j \in \S_i, j \neq k} \sum_{t \ge 1} x^i_{kt} q^i(k,j) f^i_k(t) & = &  \sum_{j \in \S_i, j \neq k} \sum_{t\ge 1} x^i_{jt} q^i(j,k) f^i_j(t) & \forall k \in \S_i\\
x^i_{kt} & \ge & 0 & \forall i, k, t\end{array}\]
\end{minipage}}
}
\caption{\label{fig:monobasic}The linear program {\sc (Whittle)}.}
\end{figure*}

This LP has infinite size, and we will fix that aspect in this section.
In particular, we now show that the LP has polynomial size when the
$f^i_k$ are piece-wise linear with poly-size specification.

\begin{define}
\label{def:break}
Given $i$, $k \in S_i$, $f^i_k(t)$ is specified as the piece-wise linear
function that passes through breakpoints $(t_1=1,f^i_k(1)),
(t_2,f^i_k(t_2)), \ldots, (t_m, f^i_k(t_m))$. Denote the set
$\{t_1,t_2,\ldots,t_m\}$ as $\W^i_k$. Therefore, for two consecutive
points $t_1, t_2 \in \W^i_k$ with $t_1<t_2$, the function $f^i_k$ is
specified at $t_1$ and $t_2$. For $t \in (t_1,t_2)$, we have
$f^i_k(t)=((t_2-t)f^i_k(t_1)+(t-t_1)f^i_k(t_2))/(t_2-t_1)$. For $t \ge
t_m$, we have $f^i_k(t) = f^i_k(t_m)$. We assume that $\W^i_k$ has
poly-size specification.
\end{define}

Consider the dual of the above relaxation. The
first constraint has multiplier $\lambda$, the second set of constraints
have multipliers $h_i$, and the final equality constraints have
multipliers $p^i_k$.  For notational convenience, define:
{\small
\begin{equation}
\label{eq:delta}
 \Delta P^i_k =   \sum_{j \in \S_i, j \neq k} \left(q^i(k,j) (p^i_j - p^i_k) \right)
 \end{equation}
} 
Note that $\Delta P^i_k$ is a variable that depends on the dual variables $p^i_*$. We obtain the following dual.
{\small \[ \mbox{Minimize } \ \ \ \  \lambda + \sum_{i=1}^n h_i  \]
\[\begin{array}{rcll}
\lambda + t h_i & \ge & r^i_k + f^i_k(t) \Delta P^i_k& \forall i, k \in \S_i, t \ge 1 \\ 
\lambda, h_i & \ge & 0 & \forall i
\end{array} \]}

Since $f^i_k(t)$ is
piece-wise linear, for two consecutive break-points $t_1 < t_2$ in
$\W^i_k$, the constraint $\lambda + t h_i \ge r^i_k + f^i_k(t) \Delta
P^i_k$ is true for all $t \in [t_1,t_2]$ iff it is true at $t_1$ and at
$t_2$.  This means that the constraints for $t \notin \W^i_k$ are
redundant.  Therefore, the above dual is equivalent to the the one
presented in Figure~\ref{fig:w}, which we denote ({\sc Whittle-Dual}).

\begin{figure*}[htbp]
\centerline{\framebox{\small
\begin{minipage}{6in}
\[ \mbox{Minimize } \ \ \ \  \lambda + \sum_{i=1}^n h_i \qquad \qquad \mbox{\sc (Whittle-Dual)} \]
\[\begin{array}{rcll}
\lambda + t h_i & \ge & r^i_k + f^i_k(t) \Delta P^i_k & \forall i, k \in \S_i, t \in \W^i_k \\ 
\lambda, h_i & \ge & 0 & \forall i
\end{array} \]
\end{minipage}
}}
\caption{\label{fig:w}The polynomial size dual of Whittle's LP, which we denote {\sc (Whittle-Dual)}.}
\end{figure*}

Taking the dual of the above program, we finally obtain a polynomial size relaxation for {\sc Monotone} bandits. Since this poly-size LP only differs from ({\sc Whittle}) in restricting $t$ to lie in the relevant set $\W^i_k$, and since it will not be explicitly needed in the remaining discussion, we omit writing it explicitly.

\subsection{The Balanced Linear Program}
\label{sec:lp}
\label{sec:balance}
We do not solve Whittle's relaxation. Instead, we solve the modification
of {\sc (Whittle-Dual)} from Figure~\ref{fig:w}, which we denote {\sc
  (Balance)}. This is shown in Figure~\ref{fig:bal}. The additional
constraint in {\sc (Balance)} (as in {\sc Feedback MAB}) is the constraint $\lambda = \sum_{i=1}^n
h_i$.

\begin{figure*}[htbp]
\centerline{\framebox{\small
\begin{minipage}{6in}
\[ \mbox{Minimize } \ \ \ \  \lambda + \sum_{i=1}^n h_i \qquad \qquad \mbox{\sc (Balance)} \]
\[\begin{array}{rcll}
\lambda + t h_i & \ge & r^i_k + f^i_k(t) \Delta P^i_k & \forall i, k \in \S_i, t \in \W^i_k \\ 
\lambda & = & \sum_{i=1}^n h_i & \\
\lambda, h_i & \ge & 0 & \forall i
\end{array} \]
\end{minipage}
}}
\caption{\label{fig:bal}The dual linear program {\sc (Balance)} for {\sc
    Monotone} MAB.}
\end{figure*}

The primal linear program corresponding to {\sc (Balance)} is the
following (where we place an unconstrained multiplier $\omega$ to the
final constraint of {\sc (Balance)}): 

{\small
\[ \mbox{Maximize       } \ \ \ \  \sum_{i=1}^n\sum_{k \in \S_i} \sum_{t \in \W^i_k} r^i_k x^i_{kt} \qquad \qquad \mbox{\sc (Primal-Balance)} \] 
\[ \begin{array}{rcll}
\sum_{i=1}^n \sum_{k \in \S_i} \sum_{t \in \W^i_k} x^i_{kt} & \le & 1 - \omega & \\
\sum_{k \in \S_i} \sum_{t \in \W^i_k} t x^i_{kt} & \le & 1 + \omega & \forall i \\
\sum_{j \neq k} \sum_{t \in \W^i_k} x^i_{kt} q^i(k,j) f^i_k(t) &= & \sum_{j \neq k} \sum_{t \in \W^i_j} x^i_{jt} q^i(j,k) f^i_j(t) & \forall i, k\\
x^i_{kt} & \ge & 0 & \forall i, k, t\end{array}\]}

The first step of the algorithm is to solve the linear program {\sc (Balance)}. Clearly the value of this LP is at least $OPT$. 
We now show the following properties of the optimal solution to {\sc (Balance)} using complementary slackness conditions between {\sc (Balance)} and {\sc (Primal-Balance}).  {\em From now on, we only deal with the optimal solutions to the above programs, so all variables correspond to the optimal setting. }

\begin{lemma}
\label{lem:opt}
Recall that $OPT$ is the optimal value to {\sc (Whittle)}. Since any feasible solution to {\sc (Balance)} is feasible to {\sc (Whittle-Dual)}, in the optimal solution to {\sc (Balance)}, $\lambda = \sum_{i=1}^n h_i \ge OPT/2$.
\end{lemma}

The next lemma is the crux of the analysis, where for any arm being played in any state, we use complementary slackness to explicitly relate the dual variables to the reward obtained. Note that unlike the analyses of primal-dual algorithms, our proof needs to use both the {\em exact} primal as well as dual complementary slackness conditions. This aspect requires us to actually solve the dual  optimally.

\begin{lemma}
\label{lem:struct1}
One of the following is true for the optimal solution to {\sc (Balance)}: Either there is a trivial $2$-approximation by repeatedly playing the same arm; or for {\em every} arm $i$ with $h_i > 0$ and for {\em every} state $k \in \S_i$, there exists $t \in \W^i_k$ such that the following LP constraint is tight with equality. 
\begin{equation} \label{tight}
\lambda + t h_i  \ge  r^i_k + f^i_k(t) \Delta P^i_k
\end{equation}
\end{lemma}
\begin{proof}
Note that if $\omega \le -1$ or $\omega \ge 1$, then the values of {\sc (Primal-Balance)} is 0, but the optimal value of  {\sc (Primal-Balance)} is at least $OPT > 0$. Thus, in the optimal solution to {\sc (Primal-Balance)}, $\omega \in (-1,1)$. 

The optimal solutions to {\sc (Balance)} and {\sc (Primal-Balance)} satisfy the following complementary slackness conditions (recall from above that $\omega > -1$ so that $1+\omega > 0$):
\begin{equation} \label{sl1}
h_i > 0 \qquad \Rightarrow \qquad \sum_{k \in \S_i} \sum_{t \in \W^i_k} t x^i_{kt} = 1+\omega>0
\end{equation}
\begin{equation} \label{sl2}
\lambda + t h_i  > r^i_k + f^i_k(t) \Delta P^i_k \qquad \Rightarrow \qquad x^i_{kt} = 0
\end{equation}

Suppose that for some $i$ such that $h_i>0$, and for some $k \in \S_i$, we have  $\lambda + t h_i  > r^i_k + f^i_k(t) \Delta P^i_k$ 
 for every $t \in \W^i_k$. By condition (\ref{sl2}), $x^i_{kt} = 0$ $\forall t \in \W^i_k$, which trivially implies that $x^i_{kt}f^i_k(t)=0$ $\forall t \in \W^i_k$.

Now, given that for a certain arm $i$ and state $k$, $x^i_{kt}f^i_k(t)=0$ $\forall t$. Therefore, in the following constraint in {\sc (Primal-Balance)}: 
$$\sum_{ l \neq k} \sum_{t \in \W^i_k} x^i_{kt} f^i_k(t)q^i(k,l)  =  \sum_{ l \neq k} \sum_{t \in \W^i_l} x^i_{lt}f^i_l(t) q^i(l,k)$$
the LHS is zero because $x^i_{kt}f^i_k(t)=0$, which means the RHS is zero. Since all variables are non-negative, this implies that for any $j \in \S_i$ with $q^i(j,k) > 0$, we have  $x^i_{jt}f^i_j(t)=0$ for all $t \in \W^i_j$. 

Recall (from Section~\ref{sec:recover}) that we assumed the graph  on the states with edges from $j$ to $k$ if $q^i(j,k) > 0$ is strongly connected. Therefore,  by repeating the above argument, we get $\forall j, t \in \W^i_j$, $x^i_{jt}f^i_j(t)=0$.

By Condition (\ref{sl1}), since $h_i>0$, there exists $j\in \S_i$ and $t \in \W^i_j$, such that $x^i_{jt}>0$ (or else the sum in Condition (\ref{sl1}) is zero). By what we proved in the previous paragraph, this implies that $f^i_j(t)=0$, which implies that $f^i_j(1)=0$ by the {\sc Monotone} property. Since $x^i_{jt}>0$, using Condition (\ref{sl2}) and plugging in $f^i_j(t)=0$, we get $\lambda + th_i = r^i_j$. Moreover, by plugging in $f^i_j(1)=0$ into the $t=1$ constraint of {\sc (Balance)}, we get $\lambda+h_i \geq r^i_j$. These two facts imply that $\lambda+h_i=r^i_j$.  The above implies that the policy that starts with arm $i$ in state $j$ and always plays this arm obtains per-step reward $\lambda+h_i > OPT/2$. 
\end{proof}

In the remaining discussion, we assume that the above lemma does not find an arm $i$ that yields reward at least $OPT/2$. This means that $\forall i,k$, there exists some $t \in \W^i_k$ that makes Inequality (\ref{tight}) tight.

\begin{lemma}
\label{lem:struct2}
For any arm $i$ such that $h_i > 0$,  and state $k \in \S_i$, if $ \Delta P^i_k < 0$, then:
$$\lambda +  h_i  =  r^i_k + f^i_k(1) \Delta P^i_k$$
\end{lemma}
\begin{proof}
By Lemma~\ref{lem:struct1} and our assumption above, Inequality (\ref{tight}) in Lemma~\ref{lem:struct1} is tight for some $t \in \W^i_k$. If it is not tight for $t=1$, then since $f^i_k(t)$ is non-decreasing in $t$ and since $\Delta P^i_k < 0$, it will not be tight for any $t$. Thus, we have a contradiction.
\end{proof}

\subsection{The {\sc BalancedIndex} Policy}
\label{sec:monoindex}
Start with the optimal solution to {\sc (Balance)}. First throw away the arms for which $h_i=0$. By Lemma \ref{lem:opt}, for the remaining arms, $\sum_i h_i \ge OPT/2$. Define the following quantities for each of these arms.

\begin{define} 
\label{def1}
For each $i$ ($h_i>0$ by assumption) and state $k \in \S_i$, let $t^i_k$ be the smallest value of $t \in \W^i_k$ for which $\lambda + t h_i  =  r^i_k + f^i_k(t) \Delta P^i_k$ in the optimal solution to {\sc (Balance)}.  By Lemma~\ref{lem:struct1}, $t^i_k$ is well-defined for every $k \in \S_i$.
\end{define}

\renewcommand{\G}{\mathcal{G}}
\renewcommand{\I}{\mathcal{I}}

\begin{define}
\label{def:pol}
For arm $i$, partition the states $\S_i$ into states $\G_i, \I_i$ as follows: 
\begin{enumerate}
\item $k \in \G_i$ if  $ \Delta P^i_k < 0$. (By Lemma~\ref{lem:struct2}, $t^i_k =1$.)
\item $k \in \I_i$ if $\Delta P^i_k \ge 0$.
\end{enumerate}
\end{define}
With the notation above, the policy is now presented in Figure~\ref{fig:restless}. In this policy, if arm $i$ has been in state $k \in \I_i$ for less than $t^i_k$ steps, it is defined to be ``not ready" for play. Once it has waited for $t^i_k$ steps, it becomes ``ready" and can be played. Moreover, if arm $i$ moves to a state in $k \in \G_i$, it is continuously played until it moves to a state in $\I_i$.

\medskip
\begin{figure}[htbp]
\centerline{\framebox{\small
\begin{minipage}{4.5in}
{\bf {\sc BalancedIndex} Policy}
\begin{tabbing}
1. \= {\bf Exploit:} Some arm $i$ moves to a state $k \in \G_i$: \\ 
\> (a) \ \ \= Play this arm exclusively as long as it remains   in a state in $\G_i$. \\
\> (b) \> {\bf Goto} step (2). \\
2. \> {\bf Explore:}  \\
\> (a) \> Play any ``ready" arm $i$ in state $k \in \I_i$. \\ 
\> \> \ \  (If no arm is ``ready", do not play at this step.) \\
\> (b) \> {\bf If} the arm moves to state in $\G_i$:   {\bf goto} Step (1), {\bf else}  {\bf goto} Step (2a).
\end{tabbing}
\end{minipage}
}}
\caption{\label{fig:restless} The {\sc BalancedIndex} Policy for {\sc
    Monotone} MAB.}
\end{figure}

Intuitively, the states in $\G_i$ are the ``exploitation" or ``good"
states. 
On the contrary, the states in $\I_i$ are ``exploration" or
``bad" states, so the policy waits until it has a high enough probability
of exiting these states before playing them. In both cases, $t_k$
corresponds to the ``recovery time" of the state, which is $1$ in a
``good" state but could be large in a ``bad" state.

\eat{Although we have not explicitly defined our policy as an index policy, we
can easily describe it using the following indices. Place a dummy arm
yielding no reward and having index $0$. When the state of arm $i$ is good
state $k \in \G_i$, the index is $2$. When the arm is in bad state $k \in
\I_i$ and not ``ready", its index is $-1$, and when it gets ``ready", the
index is $1$. Ties are broken arbitrarily. }

\paragraph{Lyapunov Function Analysis.}
\label{sec:lyap}
We use a Lyapunov (potential) function argument to show that the
policy described in Figure \ref{fig:restless} is a $2$-approximation.
Define the potential $\Phi_i$ for each arm $i$ at any time as follows.
(Recall the definition of $t^i_k$ from Definition~\ref{def1}, as well as
the quantities $\lambda$, $h_i$ from the optimal solution of {\sc
  (Balance)}.)
 
\begin{define}
If arm $i$ moved to state $k \in \S_i$ some $y$ steps ago ($y \ge 1$ by definition), the potential $\Phi_i$ is $p^i_k + h_i (\min(y,t^i_k)-1) $.
\end{define}

Therefore, whenever the arm $i$ enters state $k$, its potential is
$p^i_k$. If $k \in \I_i$, the potential then increases at rate $h_i$ for
$t^i_k-1$ steps, after which it remains fixed until the arm is played. Our
policy plays arm $i$ only if its current potential is
$p^i_k+h_i(t^i_k-1)$.

We finally complete the analysis in the following lemma. The proof
crucially uses the ``balance" property of the dual, which implies that
$\lambda=\sum_i h_i \ge OPT/2$. Let $\Phi_T$ denote the total potential,
$\sum_{i=1}^n \Phi_i$, at any step $T$ and let $R_T$ denote the total
reward accrued until that step. Define the function $\L_T = t \cdot OPT/2
- R_T - \Phi_T$. Let $\Delta R_T = R_{T+1} - R_T$ and $\Delta \Phi_T =
\Phi_{T+1} - \Phi_T$.

\begin{lemma}
\label{lem:crux2}
 $\L_T$ is a Lyapunov function. {\em i.e.}, $\E[\L_{T+1} |\L_T] \le \L_T$. Equivalently, at any step:
$$ \E[\Delta R_T + \Delta \Phi_T|R_T,\Phi_T] \ge  OPT/2$$
\end{lemma}
\begin{proof} 
  At a given step, suppose the policy does nothing, then all arms are
  ``not ready". The total increase in potential is precisely $\Delta
  \Phi_T = \sum_{i} h_i \geq OPT/2$.

  On the other hand, suppose that the policy plays arm $i$, which is
  currently in state $k$ and has been in that state for $y \ge t^i_k$
  steps. The change in reward $\Delta R_T=r^i_k$. Moreover, the current
  potential of the arm must be $\Phi_T=p_k+h_i(t^i_k-1)$.  The new
  potential follows the following distribution:
\begin{equation*}
\Phi_{T+1}=
\begin{cases} 
  p^i_j,  & \mbox{with probability } f^i_k(y)q^i(k,j) \mbox{    } \forall j \neq k \\
  p^i_k, & \mbox{with probability }  1-\sum_{j\neq k}f^i_k(y)q^i(k,j)
\end{cases}
\end{equation*}
Therefore, if arm $i$ is played, the change in potential is:
$$ \E[\Delta \Phi_T] = f^i_k(y) \sum_{j \in \S_i, j \neq k} \left(q^i(k,j) (p^i_j - p^i_k) \right) - h_i(t^i_k-1)$$
From the description of the {\sc Index} policy, $y = t^i_k = 1$ if $k \in \G_i$. Therefore, $y$ might be strictly greater than $t^i_k$ only when $k \in \I_i$. In that case $\Delta P^i_k \ge 0$ by Definition~\ref{def:pol}, so that  $f^i_j(y)\Delta P^i_k \ge f^i_j(t^i_k)\Delta P^i_k$ by the {\sc Monotone} property (since $y \ge t^i_k$). 

Therefore, for the arm $i$ being played, regardless of whether $k \in \G_i$ or $k \in \I_i$, 
\begin{eqnarray*}
\Delta R_T + \E[ \Delta \Phi_T] &= & r^i_k +  f^i_k(y) \Delta P^i_k - h_i(t^i_k-1) \\
&\ge& r^i_k +  f^i_k(t^i_k) \Delta P^i_k - h_it^i_k + h_i \\
&= & \lambda + h_i > OPT/2
\end{eqnarray*}
where the last equality follows from the definition of $t^i_k$ (Definition~\ref{def1}). Since the potentials of the arms not being played do not decrease (since all $h_l > 0$), the total change in reward plus potential is at least $OPT/2$.  This completes the proof. Refer Figure~\ref{fig:potential} for a ``picture proof" when $k \in \I_i$. 
\end{proof}

By their definition, the potentials $\Phi_T$ are bounded independent of the time horizon, by telescoping summation, the above lemma implies that $\lim_{T \rightarrow \infty} \frac{\E[R_T]}{T} \ge OPT/2$. We finally have:

\begin{theorem}
The  {\sc BalancedIndex} policy is a $2$ approximation for {\sc Monotone}  bandits.
\end{theorem}

 \renewcommand{\S}{\mathcal{S}}
 \renewcommand{\H}{\mathbf{T}}
\renewcommand{\T}{\mathbf{T}}
 \renewcommand{\p}{\mathbf{p}}

\begin{figure*}[htbp]
  \centerline{\includegraphics[width=3in]{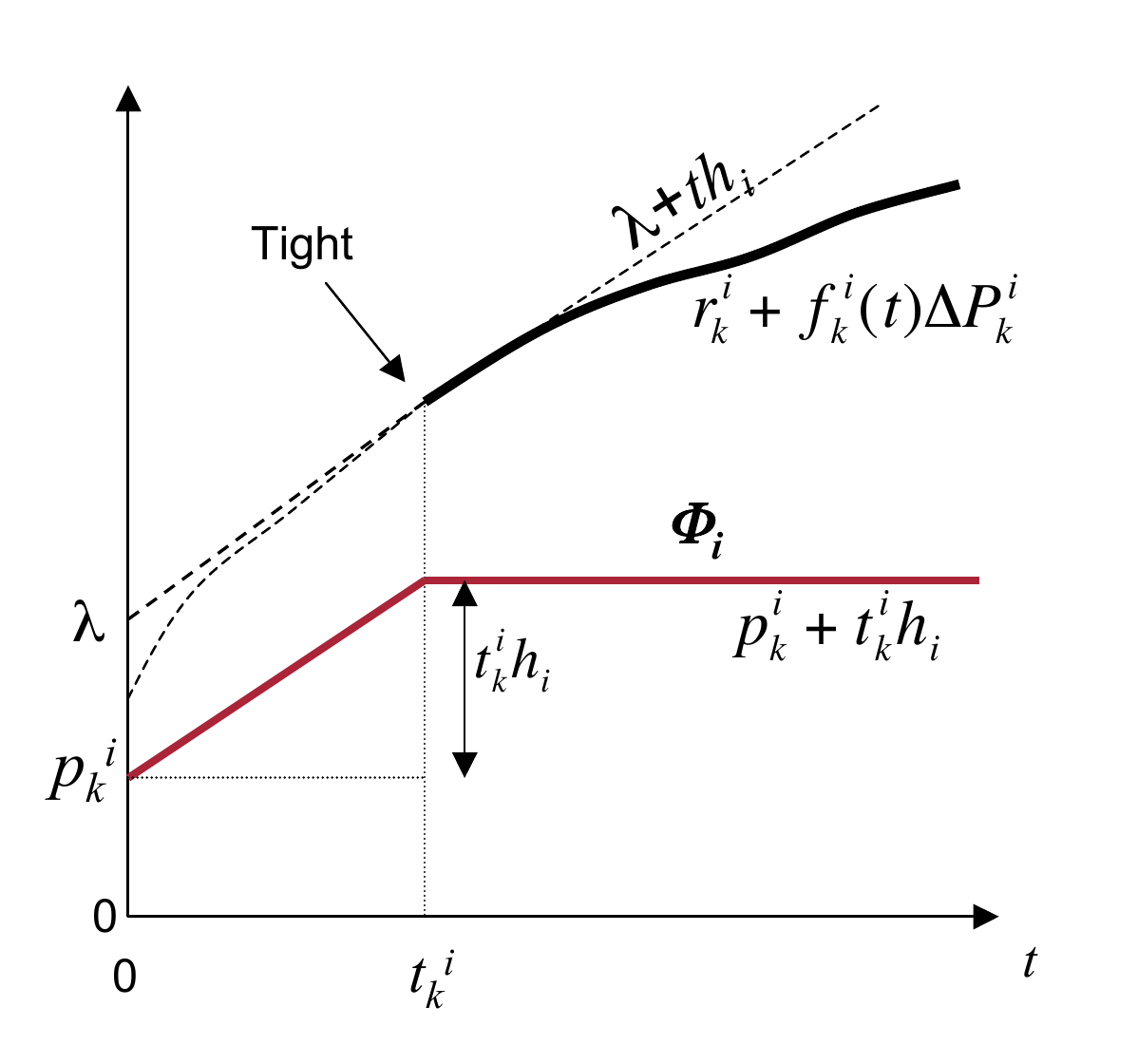}}
  \caption{\label{fig:potential} Proof of  Lemma~\ref{lem:crux2}.  The
    growth of the potential $\Phi$ is shown on the lower piece-wise linear
    function. The upper set of curves represent the LHS and RHS of the
    LP constraints for state $k \in \S_i$.  The tight point $t^i_k$ is where
    the potential switches to being constant. }
\end{figure*}

\subsection{Lower Bounds: Necessity of Monotonicity and Separability}
\label{app:hard}
We show that {\sc Monotone} bandits is NP-Hard, and that if the  {\sc Monotone} property is relaxed even slightly, the problem either has $\Omega(n)$ integrality gap for Whittle's LP, or becomes $n^{\epsilon}$-hard to approximate. 

\paragraph{Input Specification.} In the above discussion, we assumed the input to the {\sc Monotone} bandits problem is specified by polynomial size state spaces $\S_i$ for each arm; the associated matrices $q^i$, and functions $f^i_k(t)$ that are piecewise linear with poly-size specification. We can model this problem as a restless bandit problem in the sense defined in literature by replacing each state $k \in \S_i$ with exponentially many states $\{k_{t}, t \in \mathbf{Z}^+ \}$;  if the arm is not played, it transitions deterministically from state $k_t$ to $k_{t+1}$, but if played in state $k_t$, it transitions w.p. $q^i(k,j) f^i_k(t)$ to state $j_1$ for each $j \in \S_i$, and with the remaining probability transitions to $k_1$. The reduction uses exponentially many states, and is unlike the typical formulation of restless bandits that assumes the state space of each arm is poly-bounded. (The PSPACE-Hardness proofs of restless bandits~\cite{PapT} assumes poly-bounded state space as well.) We therefore need to use different NP-Hardness proofs for our compact input specifications.

\begin{theorem}
  \label{thm:hard}
  For the special case of the problem with $K=2$ states per arm and
  $n$ arms, the following are true even when the functions $f^i_k$ are piece-wise linear with poly-size specification:
  \begin{enumerate}
  \item Computing the optimal ergodic policy for {\sc Monotone} bandits is
    NP-Hard.
  \item If the {\sc Monotone} property is relaxed to allow arbitrary
    (possibly non-monotone) functions $f$, then the problem becomes
    $n^{\epsilon}$ hard to approximate unless $P=NP$.
  \end{enumerate}
\end{theorem}
\begin{proof}
  We reduce from the following periodic scheduling problem, which is shown
  to be NP-Complete in~\cite{Bhatia}: Given $n$ positive integers $l_1,
  l_2, \ldots, l_n$ such that $\sum_{i=1}^n 1/l_i \le 1$, is there an
  infinite sequence of integers $\{1,2,\ldots,n\}$ such that for every $i
  \in \{1,2,\ldots,n\}$, all consecutive occurrences of $i$ are exactly
  $l_i$ elements apart. Given an instance of this problem, for each $i \in
  \{1,2,\ldots,n\}$, we define an arm $i$ with a ``good" state $g$ and a
  ``bad" state $w$.

  For part 1, for every arm $i$, let $r^i_g = 1$, and $r^i_w = 0$. Set
  $q^i(g,w) = 1$ and $f^i_g(t) = 1$ for all $t$. Moreover, set $q^i(w,g_i)
  = 1$ and $f^i_w(t) = 0$ if $t \le 2l_i-2$ and $1$ otherwise. Suppose for
  a moment that we only have arm $i$, then the optimal policy will play the
  arm exactly $2l_i-1$ steps after it is observed to be in $w$, and the arm
  will transition to state $g$. The policy will then play the arm in state
  $g$ to obtain reward $1$, and the arm will transition back to state $w$.
  Since this policy is periodic with period $2l_i$, it yields long term
  average reward exactly $\frac{1}{2l_i}$. It is easy to see that any other
  ergodic policy of playing this arm yields strictly smaller reward per
  step. Any policy of playing {\em all} the arms therefore has total reward
  of at most $\sum_{i=1}^n \frac{1}{2l_i}$. But for any ergodic policy, the
  reward of $\sum_{i=1}^n \frac{1}{2l_i}$ is achievable {\em only if} each
  arm $i$ is played according to its individual optimal policy, which is
  twice in succession every $2l_i$ steps. But deciding whether this is
  possible is equivalent to solving the periodic scheduling problem on the
  $l_i$. Therefore, deciding whether the optimal policy to the {\sc
    Monotone} bandit problem yields reward $\sum_{i=1}^n \frac{1}{2l_i}$ is
  NP-Hard.

  For part 2, we make $w$ a trapping state with no reward. For arm $i$, set
  $q^i(g,w) = q^i(w,g) = 1$; and $f^i_g(l_i) = 0$ and $f^i_g(t) = 1$ for
  all $t \neq l_i$. Furthermore, $f^i_w(t) = 0$ for all $t$. Also set
  $r^i_g = l_i$ and $r^i_w= 0$. Therefore, for any arm $i$, any policy will
  obtain reward from this arm if and only if it chooses the start state to
  be $g$, and plays the arm periodically once every $l_i$ steps to obtain
  average reward $1$. Therefore, approximating the value of the optimal
  policy is the same as approximating the size of the largest subset of
  $\{l_1, l_2, \ldots, l_n\}$ so that this subset induces a periodic
  schedule. The NP-Hardness proof of periodic scheduling in~\cite{Bhatia}
  shows that this problem as hard as approximating the size of the largest
  subset of vertices in a graph whose induced subgraph is bipartite, which
  is $n^{\epsilon}$ hard to approximate~\cite{lund} unless $P=NP$.
\end{proof}

In the above proof, we showed that the problem becomes hard to
approximate if the transition probabilities are non-monotone. However,
that does not address the question of how far we can push our
technique. We give a negative result by showing
that Whittle's LP can have arbitrarily large gap even if the {\sc
Monotone} bandit problem is slightly generalized by preserving the
monotone nature of the transition probabilities, but removing the
additional {\em separable} structure that they should be of the form
$f^i_k(t) q^i(k,j)$.  In other words, the transition probability from
state $k$ to state $j \neq k$ if played after $t$ steps is
$q^i_{kj}(t)$ -- these are arbitrary {\em monotonically
non-decreasing} functions of $t$. We insist $\sum_{j
\neq k} q^i_{jk} (t) \le 1$ for all $k,t$ to ensure feasibility. We show that 
Whittle's LP
has $\Omega(n)$ gap for this generalization.

\begin{theorem}
 If the separability assumption on transition probabilities is
relaxed, Whittle's LP has $\Omega(n)$ gap even with $K=3$ states per arm.
\end{theorem}
\begin{proof}
The arms are all identical. Each has $3$ states, $\{g,b,a\}$. State $a$ is an
absorbing state with $0$ reward. State $g$ has reward
$1$, and state $b$ has reward $0$. The transition
probabilities are as follows: $q_{ab}(t) = q_{ag}(t)= 0$. Further, $q_{gb}(t)
= 1/2$, $q_{ga}(1) = 0$; and $q_{ga}(t) = 1/2$ for $t \ge 2$. Finally,
$q_{ba}(t) = q_{bg}(t) =  0$ for $t < 2n-1$; $q_{bg}(2n-1) = 1/2$; $q_{ba}(2n-1)= 0$; and
$q_{ba}(t) = q_{bg}(t) = 1/2$ for $t \ge 2n$.

A feasible single arm policy involves playing the arm in state $b$
after exactly $2n-1$ steps (w.p. $1/2$, the state transitions to $g$),
and continuously in state $g$ (w.p. $1/2$, the state transitions to
$b$). This policy never enters state $a$. The average rate of play is
$1/n$. The per-step reward of this policy is $\Theta(1/n)$. Whittle's LP
chooses this policy for each arm so that the total rate of play is $1$ and
the objective is $\Theta(1)$.

Now consider any feasible policy that plays at least $2$ arms. If one
of these arms is in state $g$, there is a non-zero probability that
either this arm is played after $t > 1$ steps, or the other arm in state $b$
is played after $t \ge 2n$ steps. In either case, w.p. $1/2$, the arm
enters absorbing state. Since this is an infinite horizon problem, the
above event happens w.p. $1$. Therefore, any feasible policy is
restricted to playing only one arm in the long run, and obtains reward
at most $1/n$.
\end{proof}

\section{{\sc Monotone} Bandits: Multiple Simultaneous Plays of Varying Duration}
\label{app:vary}
In this section, we extend the index policy for {\sc Monotone} bandits to handle multiple plays of varying duration.We use the same problem description as in Section~\ref{sec:recover}, except we assume there are $M \ge 1$ players, each of which can play one arm every time step. (Therefore, $M$ plays can proceed simultaneously per step.) 

Furthermore, we assume that if arm $i$ in state $k \in \S_i$ is played, this play takes $L^i_k \ge 1$ steps and during this time, this player cannot play another arm. We note that the values $L_k^i$ are fixed beforehand, and the players are aware of these values. When the player plays arm $i$ in state $k$, he/she is forced to remain on arm $i$ for $L_k^i$ steps, and he/she only receives one reward of magnitude $r^i_k$, at the beginning of this ``blocking" period.  

Suppose when the current play begins, the previous play had ended $t \ge 1$ steps ago. Then, at the end of the current play, the arm transitions to one of the states $j \neq k$ w.p.  $q^i(k,j) f^i_k(t)$, and with the remaining probability stays in state $k$. In Section~\ref{sec:recover}, we focused on the case where $M = 1$ and all $L^i_k = 1$.

Since the overall algorithm and analysis are very similar to that in Section~\ref{sec:recover}, we simply outline the differences.  First, Whittle's LP gets modified as follows:
{\small
  \[ \mbox{Maximize } \ \ \ \ \sum_{i=1}^n\sum_{k \in \S_i} \sum_{t \ge 1}
  r^i_k x^i_{kt} \qquad \qquad \mbox{\sc (Whittle)} \]
  \[ \begin{array}{rcll}
    \sum_{i=1}^n \sum_{k \in \S_i} \sum_{t \ge 1} L^i_k x^i_{kt} & \le & M & \\
    \sum_{k \in \S_i} \sum_{t \ge 1} (t + L^i_k - 1) x^i_{kt} & \le & 1 & \forall i \\
    \sum_{j \in \S_i, j \neq k} \sum_{t \ge 1} x^i_{kt} q^i(k,j) f^i_k(t) & = & \sum_{j 
\in \S_i, j \neq k} \sum_{t\ge 
1} x^i_{jt} q^i(j,k) f^i_j(t) & \forall k \in \S_i\\
    x^i_{kt} & \ge & 0 & \forall i, k \in \S_i, t \ge 1\end{array}\]} 
    In the above formulation, the first constraint merely encodes that in
expectation $M$ arms are played per step. Note that each play of arm $i$ in
state $k$ lasts $L^i_k$ steps, and the play begins with probability
$x^i_{kt}$, so that the steady state probability that arm $i$ in state $k$
is being played at any time step is $\sum_{t \ge 1} L^i_k x^i_{kt}$. Note
now that if the play begins after $t$ steps, then the arm was idle for
$t-1$ steps before this event. Therefore, the quantity $\sum_{t \ge 1} (t +
L^i_k - 1) x^i_{kt}$ would be the steady state probability that the arm $i$
is in state $k$. This summed over all $k$ must be at most $1$ for any arm
$i$. This is the second constraint. The final constraint encodes that the
rate of leaving state $k$ in steady state (LHS) must be the same as the
rate of entering state $k$ (RHS).

\subsection{Balanced Program and Complementary Slackness}
The balanced linear program is in Fig.~\ref{fig:f}. (Recall the definition of $\Delta P^i_k(t)$ from Equation~\ref{eq:delta}.) Next, Lemma~\ref{lem:struct1} gets modified as follows:
\begin{figure*}[htbp]
  \framebox{\small
    \begin{minipage}{6.0in}
      \[ \mbox{Minimize } \ \ \ \ M \lambda + \sum_{i=1}^n h_i \qquad
      \qquad \mbox{\sc (Balance)} \]
      \[\begin{array}{rcll}
        L^i_k (\lambda  + h_i) + (t - 1) h_i  & \ge & r^i_k + f^i_k(t)\Delta P^i_k(t) & \forall i, k \in \S_i, t \in \W^i_k \\ 
        M \lambda & = & \sum_{i=1}^n h_i & \\
        \lambda, h_i & \ge & 0 & \forall i
      \end{array} \]
    \end{minipage}
  }
  \caption{\label{fig:f}The linear program {\sc (Balance)} that we actually
    solve.}
\end{figure*}

\begin{lemma}
  \label{lem:struct10}
  In the optimal solution to {\sc (Balance)}, one of the following is true
  for {\em every} arm $i$ with $h_i > 0$: Either repeatedly playing the arm
  yields per-step reward at least $\lambda + h_i$; or for {\em every} state
  $k \in \S_i$, there exists $t \in \W^i_k$ such that the following LP
  constraint is tight with equality.
  \begin{equation} \label{tight10} L^i_k (\lambda + h_i) + (t - 1) h_i \ge
    r^i_k + f^i_k(t) \Delta P^i_k(t)
  \end{equation}
\end{lemma}

We next split the arms into two types:
\begin{define}
\begin{enumerate}
\item Arm $i \in U_1$  if repeatedly playing it yields average per-step reward
  at least $\lambda + h_i$. Our policy described in the next section
  favors these arms and continuously plays them.
\item Arm $i \in U_2$ if $i \notin U_1$ and $h_i > 0$. Note that for
  $i \in U_2$, $\forall k$, $\exists$ $t \in \W^i_k$ that
  makes Inequality (\ref{tight10}) tight.
\end{enumerate}
\end{define}

\begin{lemma}
  \label{lem:struct20}
  For any arm $i \in U_2$ and state $k \in \S_i$, if $\Delta P^i_k(t)< 0$, then:
$$L^i_k (\lambda  + h_i)  =  r^i_k + f^i_k(1) \Delta P^i_k(t)$$
\end{lemma}

\subsection{{\sc BalancedIndex} Policy}
\begin{define}
  \label{def10}
  For each $i \in U_2$ and state $k \in \S_i$, let $t^i_k$
  be the smallest value of $t \in \W^i_k$ for which Inequality
  (\ref{tight}) is tight.  By Lemma~\ref{lem:struct1}, $t^i_k$ is well-defined for
  every $k \in \S_i$.
\end{define}

\begin{define}
  \label{def:pol10}
  For arm $i \in U_2$, partition the states $\S_i$ into states $\G_i, \I_i$ as
  follows:
  \begin{enumerate}
  \item $k \in \G_i$ if $\Delta P^i_k(t) < 0$. (By Lemma~\ref{lem:struct20}, $t^i_k =1$.)
  \item $k \in \I_i$ if $\Delta P^i_k(t) \ge 0$.
  \end{enumerate}
\end{define}

Finally, the {\sc BalancedIndex} policy is described in Figure~\ref{fig:restless10}. Note that any arm $i \in U_2$ that is observed to be in a state in
$\G_i$ is continuously played until its state transitions into
$\I_i$. This preserves the invariant that at most $M - |U_1|$ arms $i
\in U_2$ are in states $k \in \G_i$ at any time step.

\begin{figure*}[htbp]
  \framebox{\small
    \begin{minipage}{6.0in} {\bf {\sc BalancedIndex} Policy at any time step}
      \begin{tabbing}
Continuously play $\min(M,|U_1|)$ arms in $U_1$. 
\ \ \ \ /* Execute the remaining policy only if $|U_1| < M$ */ \\ \\
{\bf Invariant:} At most $M - |U_1|$ arms $i \in U_2$ are in states $k
\in \G_i$.\\ \\
 If a player becomes free, prioritize the available arms in $U_2$ as follows: \\
 \ \  \= (a) \= {\bf Exploit:}  Choose to play an arm $i$ in state $k \in
\G_i$.  \\
\> (b) \> {\bf Explore:} If no ``good'' arm available, play any ``ready''
arm  $i$ in state $k \in \I_i$.\\
\> (c) \> {\bf Idle:} If no ``good'' or ``ready'' arm available, then
idle.  
      \end{tabbing}
    \end{minipage}
  }
  \caption{\label{fig:restless10} The {\sc Index} Policy.}
\end{figure*}

\subsection{Lyapunov Function Analysis}
 Define the potential for each arm in $U_2$ at any time as follows. 
\begin{define}
  If arm $i \in U_2$ moved to state $k \in \S_i$ some $y$ steps ago
  ($y \ge 1$ by definition), the potential is $p^i_k + h_i
  (\min(y,t^i_k)-1) $.
\end{define}

Therefore, whenever the arm $i \in U_2$ enters state $k$, its
potential is $p^i_k$.  If $k \in \I_i$, the potential then increases
at rate $h_i$ for $t^i_k-1$ steps, after which it remains fixed until
a play completes for it. When our policy decides to play arm $i \in U_2$, its
current potential is $p^i_k+h_i(t^i_k-1)$.

We finally complete the analysis in the following lemma. The proof
crucially uses the ``balance" property of the dual, which states that $M
\lambda=\sum_i h_i \ge OPT/2$. Let $\Phi_T$ denote the total potential at
any step $T$ and let $R_T$ denote the total reward accrued until that step.
Define the function $\L_T = T \cdot OPT/2 - R_T - \Phi_T$. Let $\Delta R_T
= R_{T+1} - R_T$ and $\Delta \Phi_T = \Phi_{T+1} - \Phi_T$.

\begin{lemma}
  $\L_T$ is a Lyapunov function. {\em i.e.}, $\E[\L_{T+1} |\L_T] \le \L_T$.
  Equivalently, at any step:
$$ \E[\Delta R_T + \Delta \Phi_T|R_T,\Phi_T] \ge  OPT/2$$
\end{lemma}
\begin{proof}
Arms $i \in U_1$ are played continuously and yield average per step
reward $\lambda + h_i$, so that for any such arm $i$ being played,
$\E[\Delta R_T] = \lambda + h_i$.

Next focus on arms $i \in U_2$. As before, it is easy to show that
when played, regardless of whether $k \in \G_i$ or $k \in \I_i$,
\begin{eqnarray*}
  \Delta R_i + \E[ \Delta \Phi_i] &= & r^i_k +  f^i_k(y) \sum_{j \in \S_i, j \neq k} 
\left(q^i(k,j) (p^i_j - p^i_k) 
\right) - h_i(t^i_k-1) \\
  &\ge& r^i_k +  f^i_k(t^i_k)\Delta P^i_k(t) - h_i(t^i_k -1) =  L^i_k( \lambda + h_i) 
\end{eqnarray*}
where the last equality follows from the definition of $t^i_k$
(Definition~\ref{def10}). Since the play lasts $L^i_k$ time steps, the
amortized per step change for the duration of the play, $\Delta R_T + \E[
\Delta \Phi_T]$, is equal to $\lambda + h_i$.

\medskip We finally bound the increase in reward plus potential at any
time step. At step $T$, let $S_g$ denote the arms in $U_1$ and those
in $U_2$ in states $k \in \G_i$. Let $S_r$ denote the ``ready" arms in
states $k \in \I_i$, and let $S_n$ denote the set of arms that are not
``ready".  There are two cases. If $|S_g \cup S_r| \ge M$, then some
$S_p \subseteq S_g \cup S_r$ with $|S_p| = M$ is being played.
$$\Delta R_T + \E[ \Delta \Phi_T]  \ge \sum_{i \in S_p}  (\lambda + h_i)   \ge M 
\lambda \ge OPT/2$$
Next, if $|S_g \cup S_r| < M$, then all these arms are being played.
$$\Delta R_T + \E[ \Delta \Phi_T]  = \sum_{i \in S_g  \cup S_r}  (\lambda + h_i) 
+ \sum_{i \in S_n} h_i \ge 
\sum_{i \in S_g  \cup S_r \cup S_n} h_i = \sum_i h_i \ge OPT/2$$

Since the potentials of the arms not being played do not decrease (since
all $h_l > 0$), the total change in reward plus potential is at least
$OPT/2$. 
\end{proof}

\begin{theorem}
  The {\sc BalancedIndex} policy  in Figure~\ref{fig:restless10} is a $2$ approximation for {\sc Monotone}
  bandits with multiple simultaneous plays of variable duration.
\end{theorem}

\section{{\sc Monotone} Bandits: Switching Costs}
\label{app:switch}
In several scenarios, playing an arm continuously incurs no extra
cost, but switching to a different arm incurs a closing cost for the
old arm and a setup cost for the new arm. For the applications
mentioned in Section~\ref{sec:intro}, in the context of UAV
navigation~\cite{leny}, this is the cost of moving the UAV to the new
location; or in the case of wireless channel selection, this is the
setup cost of transmitting on the new channel.

We now show a $2$-approximation for {\sc Monotone} Bandits when the
cost of switching out of arm $i$ is $c_i$ and the cost of switching
into arm $i$ is $s_i$. This cost is {\em subtracted} from the
reward. Note that the switching cost depends additively on the closing
and setup costs of the old and new arms. The remaining formulation is
the same as Section~\ref{sec:recover}.

Since the overall policy and proof are very similar to the version
without these costs, we only outline the differences. First, we define
the following variables: Let $x^i_{kt}$ denote the probability of the
event that arm $i$ in state $k$ is played after $t$ steps and this arm
was switched into from a different arm. Let $y^i_{kt}$ denote the
equivalent probability when the previous play was for the same arm.
The LP relaxation is as follows:

{\small
 \[ \mbox{Maximize       } \ \ \ \  \sum_{i=1}^n\sum_{k \in \S_i} \sum_{t \in \W^i_k} 
\left(r^i_k \left(x^i_{kt} + y^i_{kt} \right) - \left(c_i + s_i \right) x^i_{kt} \right)    
\qquad 
\qquad \mbox{\sc (LPSwitch)} \] 
  \[ \begin{array}{rcll}
    \sum_{i=1}^n \sum_{k \in \S_i} \sum_{t \in \W^i_k}   x^i_{kt} + t y^i_{kt} & \le & 
1 & \\
    \sum_{k \in \S_i} \sum_{t \in \W^i_k} t \left(x^i_{kt} + y^i_{kt} \right)& \le & 1 & 
\forall i \\
    \sum_{j \in \S_i, j \neq k} \sum_{t \in \W^i_k} (x^i_{kt} + y^i_{kt}) q^i(k,j) f^i_k(t) 
& = & \sum_{j \in \S_i, j \neq k} \sum_
{t \in \W^i_j} (x^i_{jt} + y^i_{jt}) q^i(j,k) f^i_j(t) & \forall i,k \\
    x^i_{kt}, y^i_{kt} & \ge & 0 & \forall i, k , t  \end{array}\]
The balanced dual is the following. (Recall the definition of $\Delta P^i_k(t)$ from Equation~\ref{eq:delta}.)
 \[ \mbox{Minimize } \ \ \ \ \lambda + \sum_{i=1}^n h_i \qquad
      \qquad \mbox{\sc (DualSwitch)} \]
      \[\begin{array}{rcll}
   \lambda  + t h_i  & \ge & r^i_k - c_i - s_i + f^i_k(t)\Delta P^i_k(t) & \forall i, k , t \\ 
  t (\lambda + h_i)  & \ge & r^i_k + f^i_k(t)\Delta P^i_k(t) & \forall i, k, t \\ 
        \lambda & = & \sum_{i=1}^n h_i & \\
        \lambda, h_i & \ge & 0 & \forall i
      \end{array} \]
}

The proof of the next claim follows from complementary slackness
exactly as the proof of Lemma~\ref{lem:struct1}.
\begin{lemma}
\label{lem:switch1}
  In the optimal solution to {\sc (DualSwitch)}, one of the following is true  for 
{\em every} arm $i$ with $h_i > 0$: Either repeatedly playing the arm  yields 
per-step reward at least $\lambda + h_i$; or for {\em every} state  $k \in \S_i$, 
there exists $t \in \W^i_k$ such that one of the following two LP  constraints is 
tight with equality:
\begin{enumerate}
\item $\lambda + t h_i \ge    r^i_k - c_i - s_i + f^i_k(t)\Delta P^i_k(t)$.
\item $ t (\lambda + h_i)   \ge r^i_k + f^i_k(t)\Delta P^i_k(t)$.
\end{enumerate}
\end{lemma}

Only consider arms with $h_i > 0$. The next lemma is similar to
Lemma~\ref{lem:struct2}:
\begin{lemma}
\label{lem:switch2}
  For any arm $i$ and state $k \in \S_i$, if $\Delta P^i_k(t)< 0$, then:
$ \lambda + h_i = r^i_k + f^i_k(1) \Delta P^i_k(t)$.
\end{lemma}

For arm $i$, let $t^i_k$ denote the smallest $t$ for which some dual
constraint for state $k$ (refer Lemma~\ref{lem:switch1}) is tight. The
state $k \in \S_i$ belongs to $\G_i$ if the second constraint in
Lemma~\ref{lem:switch1} is tight at $t=t^i_k$, {\em i.e.}:
\begin{equation}
\label{eqswitch}t^i_k (\lambda + h_i) = r^i_k + f^i_k(t^i_k)\Delta P^i_k(t)
\end{equation}
 By  Lemma~\ref{lem:switch2}, this includes the case where $\Delta P^i_k(t) < 0$, so that
  $t^i_k = 1$. 

Otherwise, the first constraint in Lemma~\ref{lem:switch1}
  is tight at $t = t^i_k$. This state $k$ belongs to $\I_i$, and
  becomes ``ready" after $t^i_k$ steps.  

With these definitions, the {\sc BalancedIndex} policy is as follows: Stick
  with an arm $i$ as long as its state is some $k \in \G_i$, and play
  it after waiting $t^i_k-1$ steps. Otherwise, play any ``ready"
  arm. If no ``good'' or ``ready'' arm is available, then idle.

\begin{theorem}
  The {\sc BalancedIndex} policy is a $2$-approximation for {\sc Monotone}
  bandits with switching costs.
\end{theorem}
\begin{proof}
 The definitions of the potentials and proof are the same as
 Lemma~\ref{lem:crux2}. The only difference is that the potential of
 state $k \in \G_i$ is defined to be fixed at $p^i_k$. Whenever the
 player sticks to arm $i$ in state $k \in \G_i$ and plays it after
 waiting $t^i_k-1$ steps, the reward plus change in potential
 amortized over the $t^i_k$ steps (waiting plus playing) is exactly
 $\lambda + h_i$ by Eq. (\ref{eqswitch}). The rest of the proof is the
 same as before.
\end{proof}

\section{ {\sc Feedback MAB} with Observation Costs}
\label{app:probe}
In wireless channel scheduling, the state of a channel can be accurately
determined by sending probe packets that consume energy. However, data
transmission at high bit-rate yields only delayed feedback about channel
quality. This aspect can be modeled by decoupling observation about the
state of the arm via probing, from the process of utilizing or playing the
arm to gather reward (data transmission).  We model this as a variant of
the {\sc Feedback} MAB problem, where at any step, $M$ arms can be played
without observing its state, and the reward of the underlying state is
deposited in a bank. Further, any arm can be probed by paying a cost to
determine its underlying state, and multiple such probes are allowed per
step.  The goal is to maximize the difference between the time average
reward and probing cost. A version of the probe problem was first proposed
in a preliminary draft of~\cite{GuhaMS06}.

Formally, we consider the following variant of the {\sc Feedback MAB}
problem.   As before, the
underlying $2$-state Markov chain (on states $\{g,b\})$) corresponding
to an arm evolves irrespective of whether the arm is played or
not. When arm $i$ is played, a reward of $r_i$ or $0$ (depending on
whether the underlying state is $g$ or $b$ respectively) is deposited
into a bank.  Unlike the {\sc Feedback} MAB problem, the player does
not get to know the reward value or the state of the arm.  However,
during the end of any time step, the player can probe any arm $i$ by
paying cost $c_i$ to observe its underlying state. We assume that the
probes are at the end of a time step, and the state evolves between
the probe and the start of the next time step. More than one arm can
be probed and observed any time step, but at most $M$ arms can be
played, and the plays are of unit duration. The goal as before is to
maximize the infinite horizon time-average difference between the
reward obtained from playing the arms and the probing cost spent.
Denote the difference between the reward and the probing cost as the
``value" of the policy.

Though the probe version is not a {\sc Monotone} bandit problem, we show
that the above techniques can indeed be used to construct a policy
which yields a $2+\epsilon$-approximation for any fixed $\epsilon>0$.

\subsection{LP Formulation}
Let $OPT$ denote the value of the optimal policy. The following is now
an LP relaxation for the optimal policy. Let $x^i_{gt}$
(resp. $x^i_{bt}$) denote the probability that arm $i$ was last
observed to be in state $g$ (resp.  $b$) $t$ time steps ago and
played at the current time step. Let $z^i_{gt} $ (resp. $z^i_{bt}$)
denote the probability that arm $i$ was last observed to be in state
$g$ (resp. $b$) $t$ steps ago and is probed at the current time
step. The probes are at the end of a time step, and the state evolves
between the probe and the start of the next time step. The LP
formulation is as follows, as before the LP can be solved upto a
$1+\epsilon$ factor.

{\small
\[ \mbox{Maximize} \ \ \sum_{i=1}^n  \sum_{t \ge 1} \left( r_i (u_{it}
  x^i_{gt}+v_{it}x^i_{bt})  - c_i (z^i_{gt} + z^i_{bt}) \right) \]
\[\begin{array}{rcll}
  \sum_{i=1}^n \sum_{t \ge 1} x^i_{gt}+x^i_{bt} & \le & M &\\
  \sum_{t \ge 1} t (z^i_{gt} + z^i_{bt}) & \le &1  & \forall i \\
 x^i_{st} & \le & \sum_{l \ge t}  z^i_{sl} & \forall i, t \ge 1, s \in \{g,b\}\\
\sum_{t \ge 1} (1-u_{it}) z^i_{gt} & = & \sum_{t \ge 1} v_{it} z^i_{bt} & \forall i \\
  x^i_{st},z^i_{st}  &\ge& 0&  \forall i,t \ge 1, s \in \{g,b\} \\ 
\end{array} \]}

The dual assigns a variable $\phi^i_{st} \ge 0$ for each arm $i$,
state $s \in \{g,b\}$, and last observed time $t \ge 1$. It further
assigns variables $h_i, p_i \ge 0$ per arm $i$, and $\lambda \ge 0$
globally. Let $R^i_{st}$ be the expected reward of playing arm $i$ in
state $s$ when last observed time is $t$. ($R^i_ {gt}=r_iu_{it}$,
$R^i_{gt}=r_iv_{it}$.) The balanced dual is as follows:

{\small
\[ \mbox{Minimize } M \lambda + \sum_i h_i \]
\[ \begin{array}{rcll}
 \lambda + \phi^i_{st} & \ge &R^i_{st} & \forall i, t \ge 1, s \in \{g,b\} \\
 th_i & \ge & -c_i - (1-u_{it}) p_i + \sum_{l \le t} \phi^i_{gl} & \forall i, t \ge 1 \\
 th_i & \ge & -c_i + v_{it} p_i + \sum_{l \le t} \phi^i_{bl} & \forall i, t \ge 1\\
M  \lambda &=& \sum_i h_i & \\
  \lambda, h_i, p_i, \phi^i_{st} & \ge & 0 & \forall i. s\in \{g,b\}
\end{array} \]}

We omit explicitly writing the corresponding primal. Note now that in
the dual optimal solution, $\phi^i_{st} = \max(0,R^i_{st} - \lambda)$,
$s \in \{g,b\}$. (This is the smallest value of $\phi^i_{st}$
satisfying the first constraint, and whenever we reduce $\phi^i_{st}$,
we preserve the latter constraints while possibly reducing $h_i$.)
Moreover, we have the following complementary slackness conditions:

\begin{enumerate}
\item $h_i > 0 \Rightarrow \sum_{t \ge 1} t (z^i_{gt} + z^i_{bt}) = 1 +
  \omega > 0$.
\item $z^i_{gt} > 0 \Rightarrow th_i  = -c_i - (1-u_{it}) p_i+ \sum_{l \le t} \phi^i_
{gl}$.
\item $z^i_{bt} > 0  \Rightarrow th_i  = -c_i + v_{it} p_i+ \sum_{l \le t} \phi^i_{bl}
$
\end{enumerate}

\begin{lemma}
  \label{lem:mm}
  Focus only on arms for which $h_i > 0$. For these arms, we have the   
following.
  \begin{enumerate}
  \item For at least one $t \ge 1$, $z^i_{gt} > 0$, and similarly, for some 
(possibly different) $t$, 
    $z^i_{bt} > 0$.
  \item  Let $d_i = \min \{t \ge 1,z^i_{bt} > 0\}$, then $d_i h_i = -
 c_i +  v_{id_i} p_i + \sum_{l \le d_i}\phi^i_{bt}$. Further, define $m_i=|\{ \phi^i_
{bl}>0: l\le d_i \}|$, then
$\phi^i_{bl}>0$ for $d_i-m_i+1 \le l \le d_i$ and $\phi^i_{bl}=0$ for
$l \le d_i-m_i$.
  \item Let $e_i = \min\{t \ge 1, z^i_{gt} > 0\}$. Then, for all $t
  \le e_i$, $\lambda + \phi^i_{gt} = R^i_{gt}$. Moreover,
  $e_i(\lambda+h_i) = \sum_{t \le e_i} R^i_{gt}-c_i - (1-u_{ie_i})
  p_i$.
  \end{enumerate}
\end{lemma}
\begin{proof}
  For part (1), by complementary slackness and using $h_i > 0$, we
  have $\sum_{t \ge 1} t (z^i_{gt} + z^i_{bt}) > 0$. But if
  $z^i_{gt}>0$ for some $t$, then by $\sum_{t \ge 1} (1-u_{it})
  z^i_{gt} = \sum_{t \ge 1} v_{it} z^i_{bt}$, we have $z^i_{bt}>0$ for
  some (possibly different) $t$. The reverse holds as well.

Part (2) follows by complementary slackness on $z^i_{bd_i} > 0$. The
second part follows from the fact that $\phi^i_{bl}$ is non-decreasing
since $R^i_{bl}$ is non-decreasing.
  
For part (3), since $z^i_{be_i} > 0$, by complementary slackness,
$e_ih_i = -c_i - (1-u_{ie_i}) p_i + \sum_{t \le e_i}
\phi^i_{gt}$. Note that $\phi^i_{gt} = \max(0,R^i_{gt} - \lambda)$. If
$e_i=1$, then since the LHS is positive, it must be that
$\phi^i_{gt}>0$, which implies that $\phi^i_{gt}=R^i_{gt} -
\lambda$. If $e_i>1$, then we subtract $(e_i-1)h_i \ge -c_i -
(1-u_{i(e_i-1)}) p_i + \sum_{t \le e_i-1} \phi^i_{gt}$ from the
equality and get $h_i \le (u_{ie_i} - u_{i{e_i-1}})p_i +
\phi^i_{ge_i}$. The LHS is positive and the first term of the RHS is
negative, so $\phi^i_{ge_i}>0$. Since $\phi^i_{gt}$ by the above
formula is non-increasing, $\phi^i_{gt}>0$ $\forall t\le e_i$. This in
turn implies that $\phi^i_{gt} = R^i_{gt} - \lambda$ for all $t\le
e_i$. Substituting this back into the equality yields the second
result.
\end{proof}

\subsection{Index Policy}
Let the set of arms with $h_i > 0$ be $S$, we ignore all arms except
those in $S$. The policy uses the parameters $e_i$, $d_i$ and $m_i$
defined in Lemma~\ref{lem:mm}, If arm $i$ was observed to be in state
$b$, we denote it ``not ready" for the next $d_i-m_i$ steps, and
denote it to be ``ready" at the end of the $(d_i-m_i)^{th}$ step.

 \begin{figure*}[htbp]
   \centerline{\framebox{\small
       \begin{minipage}{6.0in}
         \begin{tabbing}
 {\bf Constraint: } \ \ \= The policy keeps initiating ``Stage
           1" with ready arms  if less than $M$ arms are in either
           stages. \\ \\
        {\bf Stage 1:} \>/* Arm $i$ is last observed to be $b$ but has turned 
``ready" */ \\
           \> 1. \= {\bf Try.} Play the arm for the next $m_i$ steps. \\ 
           \> 2. \> {\bf Probe} the arm at the end of $m_i^{th}$ step. \\
           \> \>   {\bf If} state is $g$, {\bf then} go to Stage 2
           for that arm. \\ \\
          {\bf Stage 2:} \> /* Arm $i$ just observed to be in state $g$ */ \\
	   \> 1. \= {\bf Exploit.} Play the arm for the next $e_i$ steps. \\ 
           \> 2. \> {\bf Probe} the arm at the end of $e_i^{th}$ step. \\
           \> \>   {\bf If} state is $g$ {\bf then} go to 
           Step (1) of Stage (2).
         \end{tabbing}
       \end{minipage}
     }}
   \caption{\label{fig:probe2} The policy for {\sc Feedback} MAB with 
observations.}
 \end{figure*}

\begin{theorem}
  The policy  in Fig.~\ref{fig:probe2} is a $2+\epsilon$ approximation to {\sc Feedback MAB} with observation costs.
\end{theorem}
\begin{proof}
Let $OPT$ denote the $1+\epsilon$ approximate LP solution. Recall that $
\phi^i_{st} = \max(0,R^i_{st} -
\lambda)$, $s \in \{g,b\}$.

Define the following potentials for each arm $i$. If it was last
observed to be in state $b$ some $t$ steps ago, define its potential
to be $(\min(t,d_i-m_i))h_i$; if it was last observed in state $g$,
define its potential to be $p_i$. We show that  the
time-average expected value (reward minus cost) plus change in
potential per step is at least $\min(M\lambda, \sum_{i \in S }h_i )
\geq OPT/2$. Since the potentials are bounded, this proves a
2-approximation.

Each ready arm in Stage 1 is played for $m_i$ steps and probed at the
end of the $m_i^{th}$ step. Suppose that the arm was last observed to
be in state $b$ some $t$ steps ago. The total expected value is
$-c_i + \sum_{l=t}^{t+m_i-1}R^i_{bl}$, which is at least $-c_i +
\sum_{l=d_i-m_i+1}^{d_i} R^i_{bl}$ since $R^i_{bl}=r_iv_{il}$ is
non-decreasing in $l$. The expected change in potential is
$v_{i(t+m_i-1)}p_i - (d_i-m_i)h_i$, since the arm loses the potential
build up of $(d_i-m_i)h_i$ while it was not ready, and has a
probability of $v_{i(t+m_i-1)}$ of becoming good. This is at least
$v_{id_i}p_i - (d_i-m_i)h_i$ since by definition, $t \ge
d_i-m_i+1$. After $m_i$ steps,
the total expected value plus change in potential is at least $-c_i +
\sum_{l=d_i-m_i+1}^{d_i} R^i_{bl} + v_{id_i}p_i - (d_i-m_i)h_i \ge m_i h_i+
\sum_{l=d_i-m_i+1}^{d_i} (R^i_{bl} - \phi^i_{bl})$. The inequality follows by
Lemma~\ref{lem:mm} Part (2). Since $R^i_{bl} - \phi^i_{bl} = \lambda$
for $d_i - m_i + 1 \le l \le d_i$, the total expected change in value
plus potential is $m_i(\lambda+h_i)$. Thus, the average per
step for the duration of the plays is at least
$\lambda+h_i$. (This proof also shows that if $m_i=0$, then the
probing on the previous step does not decrease the potential.)

Similarly, each arm $i$ in Stage 2 was probed and found to be good, so
that it is exploited for $e_i$ steps and probed at the end of the
$e_i^{th}$ step. During these $e_i$ steps, the total expected value is
$\sum_{t \le e_i} R^i_{gt}-c_i$, and expected change in potential is
$- (1-u_{ie_i}) p_i$, since the arm has probability $(1-u_{ie_i})$ of
being in a bad state at the end. By Lemma~\ref{lem:mm}, Part (3), the
total expected value plus change in potential is $ \sum_{t
\le e_i} R^i_{gt}-c_i - (1-u_{ie_i}) p_i = e_i(\lambda+h_i)$, so the
average change per step is $\lambda+h_i$.

Now, if $M$ arms are currently in Stage 1 or 2, then the total value
plus change in potential for these arms is at least $M\lambda \ge
OPT/2$. If fewer than $M$ arms are in those stages, then every arm $i$
that is not in Stage 1 or Stage 2 is in state $b$ and not
``ready''. Thus, its change in potential is $h_i$. Moreover, for every
arm $j$ that is in Stage 1 or Stage 2, we also get a contribution of
at least $\lambda+h_j\ge h_j$. Summing, we get a expected value plus
change in potential of at least $\sum_{i \in S}h_i \ge OPT/2$, which completes
the proof.
\end{proof}

\section{Non-Preemptive Machine Replenishment}
\label{sec:beyond}
Finally, we show our technique of balancing provides a $2$-approximation
for an unrelated, yet classic, restless bandit
problem~\cite{book,trivedi,Shi2007}: modeling breakdown and repair of
machines. Interestingly, we also show that the Whittle index policy is an
arbitrarily poor approximation to non-preemptive machine replenishment,
and thus the technique we suggest can be significantly stronger than the
Whittle index policies.

There are $n$ independent machines whose performance
degrades with time in a Markovian fashion. This is modeled by
transitions between states yielding decreasing rewards. At any step,
any machine can be moved to a repair queue by paying a cost. The
repair process is non-preemptive, Markovian, and can work on at most
$M$ machines per time step.  A scheduling policy decides when to move
a machine to a repair queue and which machine to repair at any time
slot. The goal is to find a scheduling policy to maximize the
time-average difference between rewards and repair cost. Note that if
an arm is viewed as a machine, playing it corresponds to repairing it,
and does not yield reward. In that sense, this problem is like an inverse
of the {\sc Monotone} bandits problem. We emphasize that the repairs
are {\em non-preemptive}, which means that once a repair is started, it
cannot be stopped.

Formally, there are $n$ machines. Let $\S_i$ denote the set of active
states for machine $i$. If the state of machine $i$ is $u \in \S_i$ at time
the beginning of time $t$, the state evolves into $v \in \S_i$ at time
$t+1$ w.p.  $\p_{uv}$. The state transitions for different machines when
they are active are independent. If the state of machine $i$ is $u \in
\S_i$ during a time step, it accrues reward $r_u \ge 0$. We assume each
$S_i$ is poly-size.

During any time instant, machine $i$ in state $u \in \S_i$ can be
scheduled for maintenance by moving it to the repair queue starting
with the next time slot by paying cost $c_u$. The maintenance process
for machine $i$ takes time which is distributed as {\tt
Geometric}$(s_i)$, independent of the other machines. Therefore, if
the repair process works on machine $i$ at any time step, this repair
completes after that time step with probability $s_i$.  During the
time when the machine is in the repair queue, it yields no
reward. When the machine is in the repair queue, we denote its state
by $\kappa_i$. The maintenance process is non-preemptive, and the
server can maintain at most $M$ machines at any time. When a repair
completes, the machine $i$ returns to its ``initial active state"
$\rho_i \in \S_i $ at the beginning of the next time slot. The goal is
to design a scheduling policy so that the time- average reward minus
maintenance cost is maximized.

\medskip In related work, Munagala and Shi~\cite{Shi2007} showed using a
novel queuing analysis that when the repair process is {\em preemptive},
$M=1$, and when $\S_i = \{\rho_i,b_i\}$ for all machines $i$, and $r_{b_i}
= 0$, so that the machine is either ``active" (state $\rho_i$) or
``broken" (state $b_i$), then a simple greedy policy that is equivalent to
the Whittle index policy is a $1.51$ approximation. However, as we show
later, the Whittle index policy can be arbitrarily bad for non-preemptive
repairs since it computes indices for each machine separately. We now show
that our duality-based technique yields a $2$-approximation policy with
general $\S_i$, $M$, and non-preemptive repairs.

 \subsection{LP Formulation and Dual}
 We now present an LP bound on the optimal policy. For any policy, let
 $x_u$ denote the steady state probability that machine $i$ is in state $u$
 during a time step, and $z_u$ denote the steady state probability that the
 machine $i$ transitions from state $u \in \S_i$ to state $\kappa_i$. We
 assume the policy moves a machine to the repair queue at the beginning of
 a time slot, and that repairs finish at end of a time slot. Note that it does not 
make sense to repair a 
machine in its initial state $x_{\rho_i}$ so $z_{\rho_i}=0$.

{\small
 \[\mbox{Maximize } \sum_{i} \sum_{u \in \S_i} \left( r_u x_u - c_u z_u
 \right) \]
 \[ \begin{array}{rcll}
   \sum_i x_{\kappa_i} &\le & M &\\
   x_{\kappa_i} +  \sum_{u \in \S_i} x_u & \le &  1& \forall i \\
   \sum_{v \in \S_i, v\neq u} x_v  \p_{vu} & = & z_u + \sum_{v \in S_i, v\neq u}
x_u \p_{uv}&  \forall i, u \in 
\S_i \setminus \{\rho_i\} \\
   s_i x_{\kappa_i} +   \sum_{v \in \S_i, v\neq u} x_v  \p_{v \rho_i}& = &  \sum_{v 
\in S_i, v\neq \rho_i}x_
{\rho_i} \p_{\rho_i v} & \forall i\\
   z_u, x_u & \ge & 0 & \forall i, u \in \S_i \cup \{\kappa_i\}
 \end{array}\]
}

 The dual of the above LP assigns potentials $\phi_u$ for each state $u \in
 \S_i$. Further, it assigns a value $h_i \ge 0$ for each machine $i$, and a
 global variable $\lambda \ge 0$. We directly write the balanced dual:

{\small
 \[ \mbox{Minimize } M \lambda + \sum_i h_i\]
 \[ \begin{array}{rcll}
   \lambda + h_i & \ge & s_i \phi_{\rho_i} & \forall i \\
   h_i & \ge & r_u + \sum_{v \in \S_i} \p_{uv} (\phi_v - \phi_u) & \forall i, u \in \S_i  
\\
   \phi_u + c_u & \ge & 0 & \forall i, u \in \S_i \\
   M \lambda &=& \sum_i h_i &\\
   \lambda, h_i & \ge & 0 & \forall i
 \end{array}
 \]}

 Note that $M \lambda = \sum_i h_i \ge OPT/2$. We omit explicitly writing the
 corresponding primal formulation. Now, {\em Focus only on machines for 
which 
$h_i > 0$.} We have the following complementary slackness conditions:

 \begin{enumerate}
 \item $h_i > 0 \Rightarrow x_{\kappa_i} + \sum_{u \in \S_i} x_u\ = 1 -
   \omega > 0$
 \item $ x_u > 0 \Rightarrow h_i = r_u + \sum_{v \in \S_i} \p_{uv} (\phi_v
   - \phi_u)$.
 \item $z_u > 0 \Rightarrow \phi_u + c_u = 0$.
 \item $ x_{\kappa_i} > 0 \Rightarrow \lambda + h_i = s_i \phi_{\rho_i}$.
 
 \end{enumerate}

 \subsection{Index Policy and Analysis}

\begin{figure}[htbp]
  \centerline{\framebox{
      \begin{minipage}{6.0in}
        \begin{tabbing}
          {\bf Scheduling:} \ \  \=  Consider only machines $i$ with $h_i>0$\\
          \> \ \ \=  {\bf If} the machine in in state $u$ where $z_u > 0$, then move 
the machine to repair 
queue. \\ \\
          {\bf Repair:}  \> Service any subset $S_w$ of at most $M$ machines in 
the repair 
queue non-preemptively.\\
          \> \> {\bf w.p. } $s_i$, the service for $i \in S_w$ completes  and it moves 
to state $\rho_i$.
        \end{tabbing}
      \end{minipage}
    }}
  \caption{\label{fig2} The repair policy from our LP-duality approach}
\end{figure}

\begin{claim}
  \label{claim:silly}
Consider only machines with $h_i>0$. There are two cases:
\begin{enumerate}
\item For machines in which $z_v>0$ for some $v$, we have $x_{\kappa_i} > 
0$ so that the policy can 
only reach states $u \in \S_i$ in which $x_u +  z_u > 0$. 
\item For machines in which $z_v=0$ for all $v$, we have $x_{\kappa_i} = 0$. 
The policy will never 
repair the machine, and after a finite number of steps,, the machine will only 
visit states $u \in \S_i$ for 
which $x_u>0$. 
\end{enumerate}
\end{claim}
\begin{proof}
Adding the third and fourth constraints of the primal yields $s_ix_{\kappa_i} = 
\sum_{u\in \S_i} z_u$. If for 
some $v$, $z_v>0$, then $x_{\kappa_i}>0$, which by the fourth constraint in 
the primal implies that $x_
{\rho_i}>0$. Now, suppose that $x_v>0$, then for every state $u$ such that $
\p_{vu}>0$, the third 
constraint in the primal implies that $z_u+x_u>0$. If $z_u>0$, then the policy 
will stop at state $u$ and 
enter machine $i$ into the repair queue. If $z_u=0$, then it must be that 
$x_u>0$. Repeatedly using the 
above argument starting at $v=\rho_i$, we see that the policy will only visit 
states with $x_u + z_u>0$, 
not going beyond the first state where $z_u > 0$.

For machines in which $z_v=0$ for all $v$, conditions (3) and (4) in the primal 
imply that $\{ x_v\}$ are 
the steady state probabilities of a Markov chain with transition matrix $[ \p_{uv} 
]$. Therefore, after a finite 
number of steps, the machine will only go to states $u \in \S_i$ for which 
$x_u>0$.
\end{proof}

\begin{theorem}
  The  policy  in Fig.~\ref{fig2} is a $2$-approximation for non-preemptive 
machine
  replenishment.
\end{theorem}
\begin{proof}
  We next interpret $\phi_u$ as the potential for state $u \in \S_i$. Let
  the potential for state $\kappa_i$ be $0$. We show that in each step, the
  expected reward plus change in potential is at least $OPT/2$.

  First, when any active machine $i$ enters a state $u$ with $z_u > 0$,
  then the machine is moved to the repair queue by paying cost $c_u$. The
  potential change is $-\phi_u$, and the sum of the cost and potential
  change is $-c_u - \phi_u$. The last term is $0$ by complementary
  slackness. Therefore, moving a machine to the repair queue does not
  alter the potential.

  Next, let $S_r$ denote the set of machines in the repair queue, and let
  $S_w \subseteq S_r$ denote the subset of these machines being repaired at
  the current time. Note that if $|S_r| < M$, then $S_w = S_r$, otherwise
  $|S_w| = M$. For each machine $i \in S_w$, the repair finishes w.p. $s_i$, 
and the
  machine's potential changes by $\phi_{\rho_i}$. Therefore, the expected
  change in potential per step is $s_i \phi_{\rho_i} = \lambda + h_i$ by
  complementary slackness.

  Suppose first that $|S_w| = M$, then the net reward plus change in
  potential is at least $M(\lambda + h_i) > M \lambda \ge OPT/2$. Suppose that 
$|S_w|<M$, then 
  must have $S_w = S_r$. Note that any machine that enters a state $u$ with 
$z_u>0$ will be 
automatically moved to $S_r$ at the beginning of the time step. Using this 
along with Claim~\ref
{claim:silly}, we have that after a finite number of steps, any machine $i \notin 
S_r$ enters a state $u$ 
with $x_u > 0$. (Since we care about infinite horizon average reward, the 
finite number of steps don't 
matter.) The reward plus change
  in potential for machine $i \notin S_r$ is $r_u + \sum_{v \in \S_i} \p_{uv} 
(\phi_v
  - \phi_u) = h_i$ by complementary slackness. Therefore, the total reward
  plus change in potential is $\sum_{i \in S_r} (\lambda + h_i) + \sum_{i
    \notin S_r} h_i \ge \sum_i h_i \ge OPT/2$. Since the potentials are bounded, 
the policy is a 2-
approximation.
\end{proof}

\subsection{Gap of the Whittle Index}
We now show that the Whittle index policy is an arbitrarily poor approximation to non-preemptive machine replenishment.  Note
that in the situation shown below, Whittles index  is a $1.51$ approximation when repairs can be
preempted~\cite{Shi2007}. However, when no preemption is allowed, the policy can perform arbitrarily poorly.

\begin{theorem}
The Whittle index policy is an arbitrarily poor approximation for non-preemptive machine
replenishment even with $2$ machines and $M=1$ repairs per step.
\end{theorem}
\begin{proof}
Suppose $M=1$, there are two machines $\{1,2\}$, and $\S_i =
\{\rho_i,b_i\}$ for machines $i \in \{1,2\}$. Let $r_{\rho_i} = r_i$ and
let $r_{b_i} = 0$, so that the machine is either ``active" (state $\rho_i$)
or ``broken" (state $b_i$). Let $p_i$ denote the probability of
transitioning from state $\rho_i$ to $b_i$. Assume $c_i = 0$. Note that
playing a machine corresponds to moving it to the repair queue.

The Whittle index of a state is the largest penalty that can be charged per
maintenance step so that the optimal single machine policy will still
schedule the machine for maintenance on entering the current state. In
$2$-state machines mentioned above, the Whittle index in state $\rho_i$ is
negative, since even with penalty zero per repair step, the policy will not
schedule the machine for maintenance in the good state.  The Whittle index
for state $b_i$ is $\eta_i = s_i r_i/p_i$, since for this value of penalty,
the expected reward of $r_i/p_i$ per renewal is the same as the expected
penalty of $\eta_i/s_i$ paid for maintenance in the renewal period. 

Suppose $s_1 = 1/n^4$, $s_2 = 1$, $p_1 = 1/n$, $p_2 = 1$, $r_1 = r_2 =
1$.  If used by itself, machine $1$ yields reward $r_1 \frac{s_1}{s_1
+ p_1} \approx 1/n^3$ and machine $2$ yields reward $r_2
\frac{s_2}{s_2 + p_2} = 1$. Any reasonable policy will therefore only
maintain machine $2$ and ignore machine $1$. However, in the Whittle
index policy, when machine $1$ is broken and machine $2$ is active,
the policy decides to maintain machine $1$ (since the Whittle index,
$\eta_1$, of $b_1$ is positive and that of $\rho_2$ is negative). In
this case, machine $1$ is scheduled for repair. This repair takes
$O(n^4)$ time steps and cannot be interrupted. Moreover, since machine
2 is bad at least half the time, this ``blocking" by machine 1 will
happen with rate $O(1/2)$, so in the long run, machine 2 is almost
always broken and the Whittle index  policy obtains reward $O(1/n^3)$, while
the optimal policy obtains reward $r_2 \frac{1}{1 + 1} = 1/2$ by only
maintaining machine $2$. 
\end{proof}

\section{Open Questions}
\label{sec:structure}
Our work throws open interesting research avenues. First, can our algorithmic techniques be extended to other  subclasses of restless bandits,  for instance, the POMDP problem obtained by  generalizing {\sc Feedback} MAB to $K > 2$ states per arm? Note that unlike the $K=2 $ case considered here, the transition probability values are no longer monotone as they are  based on an underlying Markov chain. Next, can matching hardness results be shown for these problems, particularly {\sc Feedback MAB}. Finally, our analysis effectively uses  piece-wise linear Lyapunov functions. Such functions derived from LP  relaxations have also been used by Bertsimas, Gamarnik, and Tsitsiklis~\cite{gamarnik} to show stability in multi-class queuing systems. Though the  techniques and results in that work  are very different from ours, it would be  interesting to explore whether our techniques extend to multi-class queuing  problems.

\paragraph{Acknowledgment.} 
We thank Shivnath Babu, Jerome Le Ny, Ashish Goel, and Alex Slivkins
for discussions concerning parts of this work.  We also thank the 
anonymous FOCS 2007 and SODA 2009 reviewers for several helpful comments on earlier
drafts of this work.

{\small
\bibliographystyle{plain} 
}

\appendix
\section{Omitted Proofs}
\label{app:omitted}
The following proofs are deferred here because they are independent of our duality based technique, and because of their lengths, we fear that they may detract from the paper's flow. 

\subsection{Proof of Theorem~\ref{thm:no-opt-index}}
We show examples in which the myopic policy and the optimal index policy exhibit the desired gaps against the optimum.

\paragraph{Gap of the Myopic Policy.}
We now show an instance where the
myopic policy that plays the arm with the highest expected next-step
reward has gap $\Omega(n)$ with respect to the reward of the optimal
policy. The instance is an extension of the instance constructed above.
There is one ``type 1" deterministic arm with reward $r_1 = 1$. There are
$n$ i.i.d. ``type 2" arms with $r_2 = n$, $\beta = \frac{1}{2^n}$, and
$\frac{\alpha}{\alpha+\beta} = \frac{1}{n}$.

First consider the myopic policy. Any policy encounters an instant where
all the type 2 arms are in state $b$. In this case, the myopic next step
reward of any of these arms is at most $r_2 \frac{\alpha}{\alpha+\beta} <
1$, so that the policy always plays the type $1$ arm, yielding long-term
reward of $1$.

Next consider the myopic policy that ignores the type 1 arm. Such a policy
performs round-robin on the arms when it observes all of them to be in
state $b$. In this case, the probability that the arm it plays will be in
state $g$ is at least $v_{n} \ge \frac{1}{2^n}$.  Therefore, the behavior
of this policy is dominated by the following $2$-state Markov chain: The
two states are $h$ and $l$; state $h$ yields reward $n$, and state $l$,
reward $0$. The transition probabilities from $h$ to $l$ and vice-versa
are $\frac{1}{2^n}$. The long-term reward is therefore at least
$\frac{n}{2}$, which lower-bounds the reward of the optimal policy. 

\paragraph{\bf Non-optimality of Index policies.}
We will now show an instance where
there is a constant factor gap between the optimal policy and the optimal
index policy. The example has $3$ arms. Arm $1$ is deterministic with
reward $r_1 = 1$. Arms $2$ and $3$ are $i.i.d.$ with $\alpha = \beta =
0.1$ and reward in state $g$ being $r_2 = 2$.

We compute the optimal policy by value iteration~\cite{book} using a
discount factor of $\gamma = 0.99$ (to ensure the dynamic program
converges). The optimal policy always plays arms $2$ or $3$ if either was
just observed in state $g$. The decisions are complicated only if both
arms $2,3$ were last observed in state $b$. In this case, we can compactly
represent the current state by the pair $(k_1,k_2) \in \mathbf{Z^+}\times
\mathbf{Z^+}$, representing the time steps ago that arms $2,3$ were
observed in state $b$ respectively. For such a state, the policy either
plays arm $1$; or plays arms $2$ or $3$ depending on whether $k_1 > k_2$
or not. Such a policy is therefore completely characterized by the region
$\D$ on the $(k_1,k_2)$ plane where the decision is to play arm $1$; in
the remaining region, it plays arm $2$ or $3$ depending on whether $k_1 >
k_2$ or not.  For the optimal policy, we have:
$$ \D^* = \left\{(k_1,k_2) \in  \mathbf{Z^+}\times \mathbf{Z^+} \ \  |\ \  k_1 \le 4, k_2 \le 4, k_1 + k_2 \le 6 \right\}$$
In other words, the description of the optimal policy is as follows (note that it is symmetric w.r.t. arms $2,3$):

\begin{table}[htbp]
\centerline{
\begin{tabular}{|l|c|}
\hline
State $(k_1,k_2)$ & Play Arm\\
\hline
$k_1 \le 3, k_2 < k_1$ & 1 \\
$k_1 = 4, k_2 \le 2$ & 1 \\
$k_1 = 4, k_2 = 3$ & 2 \\
$k_1 = 4,k_2 > 4$ & 3 \\
$k_1 \ge 5, k_2 < k_1$ & 2 \\
\hline
\end{tabular}}
\caption{Description of the optimal policy when both arms $2,3$ were last observed to be $b$.  Note that the policy is symmetric w.r.t. arms $2,3$, and furthermore, $k_1 \neq k_2$.}
\end{table}

Note the following non-index behavior where given the state of arms $1$
and $2$, the decision to play switches between these arms depending on the
state of arm $3$. If arm $2$ was observed to be $b$ some $4$ steps ago,
then: (i) If arm $3$ was $b$ some $2$ steps ago, the policy plays arm $1$;
(ii) If arm $3$ was $b$ some $3$ steps ago, the policy plays arm $2$. To
compute the reward of this policy, we observe that it has an equivalent
description as a Markov Chain over $6$ states (these new states correspond
to groups of states in the original process). A closed form evaluation of
this chain shows that the reward of the optimal policy is $1.46218$.

We next perform this evaluation for the nearby index policies. Note that
for any index policy, the region $\D$ has to be an axis-parallel square.
The first is where the decision for $k_1 = 4, k_2 \le 2$ is to play arm
$2$, so that $\D = \{(k_1,k_2) | k_1, k_2 \le 3\}$. This policy evaluates
to an average reward $1.46104$. The next is where the decision for $k_1 =
4, k_2 = 3$ is to play arm $1$, so that $\D = \{(k_1,k_2) | k_1, k_2 \le
4\}$. This policy has reward $1.46167$. Other index policies have only
worse reward. This implies that there is a constant factor gap between the
optimal policy and the best index policy.

\begin{figure}[htbp]
\centerline{  \includegraphics[width=3in]{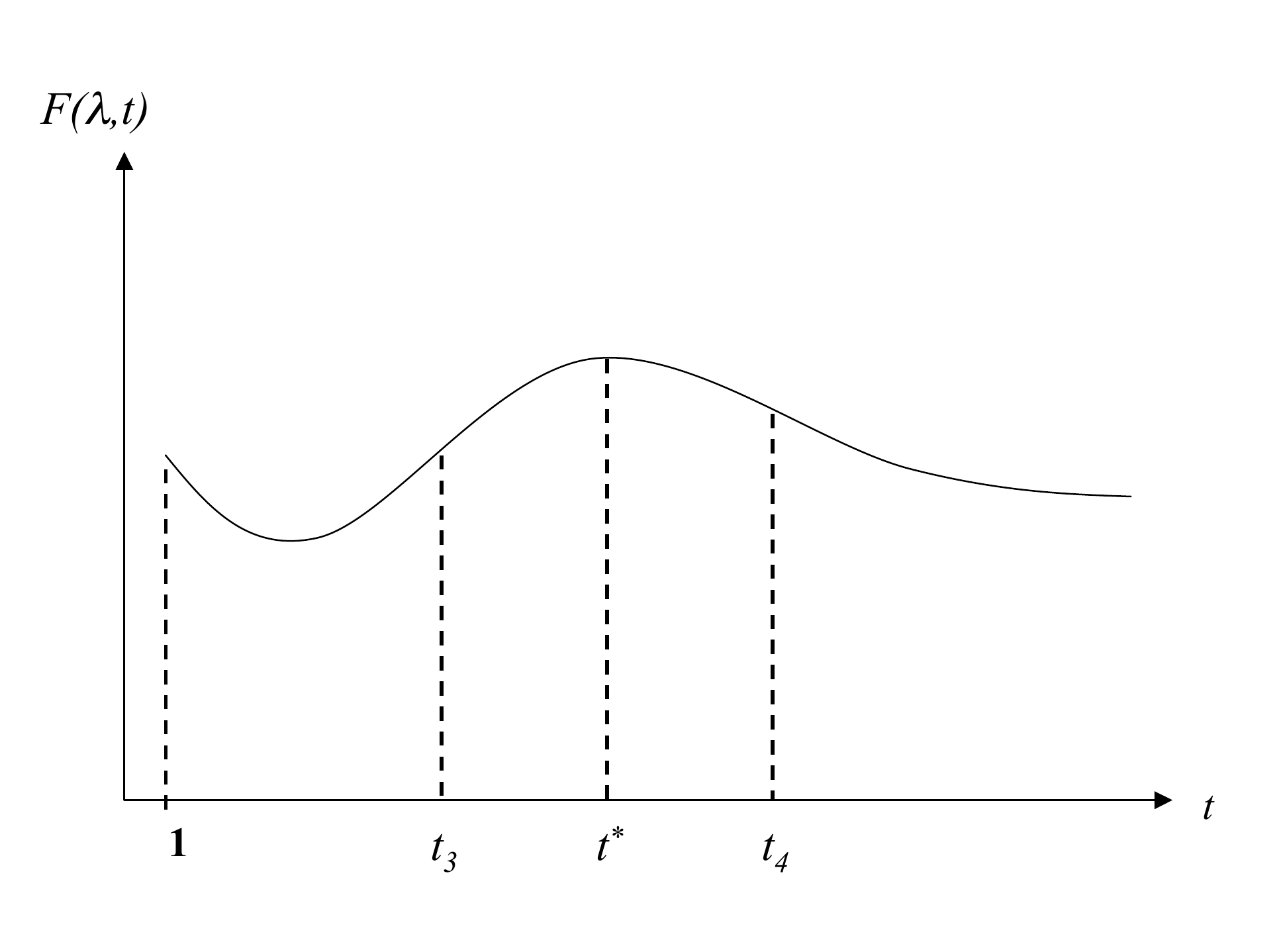}}
\caption{ \label{fig:behav} Behavior of the function $F(\lambda,t)$. }
\end{figure}

\subsection{Proof of Theorem~\ref{lem:poly}}
Since the proof focuses on a single arm $i$, we omit the subscript for the arm. For notational convenience, denote $t^* = t(\lambda)$. The expression for $H(\lambda)$ follows easily from Lemma~\ref{lem:main1}. Recall from Definition~\ref{def:rq} that $F(\lambda,t)=R(t)-\lambda Q(t)$ as the value of policy $\P(t)$.

\paragraph{Case 1:} $\lambda \ge  r \left(\frac{\alpha}{\alpha +
\beta(\alpha+\beta)} \right)$. 
Consider the subcase $\lambda \geq r$. The function $F(\lambda,t)$ is maximized by
driving the expression (which is always
non-positive) to zero. This happens when $t =\infty$. Otherwise, when $r >
\lambda $ observe (using the upper bound of $v_t$) that
$$ F(\lambda,t) = \frac{(r-\lambda) v_t - \lambda \beta}{v_t + t \beta}
\le  \frac{(r-\lambda)\frac{\alpha}{\alpha+\beta} - \lambda
  \beta}{v_t + t \beta} $$
The above is now non-positive, and it follows again that $t=\infty$ is the optimum solution.

\paragraph{Case 2:} In this case let $\lambda  =  r \left(\frac{\alpha}{\alpha +
\beta(\alpha+\beta)} \right) - \rho$ for some $\rho>0$. Rewrite the above
expression as
 
$$ F(\lambda,t) = (r - \lambda) - \beta \frac{\lambda + t (r-\lambda)}{v_t + t \beta}$$
Define the following quantities (independent of $t$):
$$ \nu = (1-\alpha - \beta) \qquad \eta =
\frac{\alpha}{\alpha+\beta}\log\frac{1}{\nu} \qquad \phi = \eta \lambda +
\frac{\alpha}{\alpha+\beta}(r-\lambda)$$
$$ \mu = \eta (r-\lambda)
\qquad \omega = \lambda \beta - \alpha \frac{r - \lambda}{\alpha+\beta} =
- \rho \frac{\alpha+\beta(\alpha+\beta)}{\alpha+\beta}$$

Observe $r-\lambda > \rho$. Note that $\phi, \mu \ge 0$. By  assumption, the value $\nu \in (\delta,1]$ has polynomial bit complexity. The same holds for $\eta,\phi,\mu$ and $\rho$. Relaxing $t$ to be a real, observe:

\[ \frac{\partial
  F(\lambda,t)}{\partial t} = \frac{\beta}{(v_t + t \beta)^2} \left((\phi
  + \mu t) \nu^t + \omega \right) \]

Since the denominator of $\partial F/\partial t$ is always non-negative, the value of  $t^*$ is either $t^* =1$, or the point where the  sign of the numerator $g(t)=(\phi +
\mu t) \nu^t + \omega $ changes from $+$ to $-$. We observe that $g(t)$ has a unique maximum at $t_3 = \frac{1}{\log (1/\nu)} - \frac{\phi}{\mu}$.  If $g(t_3)$ is
negative then the numerator of $\frac{\partial F(\lambda,t)}{\partial t}$
is always negative and the optimum solution is at $t^*=1$.

If $g(t_3)$ is positive, then it cannot change sign from $+$ to $-$ in the
range $[1,t_3)$ since it has a unique maximum. Therefore in this range
$t=1$, $t = \lfloor t_3 \rfloor$, or $t = \lceil t_3\rceil$  are the optimum solutions. 

But for $t\ge t_3$ since $g(t)$ is decreasing, $\partial F/\partial t$ changes sign
once from $+$ to $-$ as $t$ increases, and $\rightarrow 0$ as $t \rightarrow
\infty$. This behavior is illustrated in Figure~\ref{fig:behav}.  Therefore, we find a $t_4> t_3$ such that $g(t_4) < 0$, and perform binary search in
the range $[t_3,t_4]$ to find the point where $F$ is maximized.  It is easy
to compute $t_4$ with polynomial bit complexity in  the complexities
of $\nu,\eta,\phi,\mu$ and $\rho$.  We finally compare this
maximal value of $F$ to the values of $F$ at $1,\lfloor t_3 \rfloor, \lceil
t_3\rceil$. Thus we can solve $H(\lambda)$ and obtain $t^*$ in polytime.

\subsection{Proof of Theorem~\ref{thm:whittle-gap}}
\label{app:lp}
We first show the structure of the optimal solution to {\sc (Whittle)}. Using the notation from Definition~\ref{def:rq}, we have: $H_i(\lambda) = R_i(\lambda) - \lambda Q_i(\lambda)$. Let $R(\lambda) = \sum_{i=1}^n R_i(\lambda)$ and $Q(\lambda) = \sum_{i=1}^n Q_i(\lambda)$.  The following lemma shows that the optimal solution to {\sc (Whittle)} is obtained by choosing $\lambda$ such that $Q (\lambda) \approx 1$. 

\begin{lemma}
\label{thm:combine}
The optimal solution to Whittle's LP chooses a penalty $\lambda^*$  and a fraction $a \in [0,1]$, so that $a Q(\lambda_-^*) + (1-a) Q(\lambda_+^*) = 1$.  Here, $\lambda^*_- \le \lambda^* < \lambda^*_+$ with $|\lambda^*_+ - \lambda^*_-| \rightarrow 0$. The solution corresponds to a convex combination of $\P_i(t_i(\lambda^*_-))$ with weight $a$ and $\P_i(t_i(\lambda^*_+))$ with weight $1-a$ for each arm $i$.
\end{lemma}
\begin{proof}
For the optimal solution to {\sc (Whittle)}, recall that $OPT$ denote the expected reward. The expected rate of playing the arms is exactly $1$ by the LP constraint. 

When $\lambda = 0$, then $t_i(\lambda) =1$ for all $i$, implying $Q(\lambda) = n$. Similarly, when $\lambda = \lambda_{\max} \ge \max_i r_i$, $t_i(\lambda) = \infty$ for all $i$, so that $Q(\lambda) = 0$. Therefore, as $\lambda$ is increased from $0$ to $\lambda_{\max}$, there is a transition value $\lambda^*$ such that $Q(\lambda_-^*) = Q_1 \ge 1$, and $Q(\lambda^*_+) = Q_2 < 1$; furthermore, $|\lambda^*_+ - \lambda^*_-| \rightarrow 0$.  

Since the solution to {\sc (Whittle)} is feasible for {\sc LPLagrange}$(\lambda)$, we must have:
\[ R(\lambda^*_+) - \lambda^* Q_2 \ge  OPT - \lambda^*  \qquad
R(\lambda^*_-) - \lambda^* Q_1 \ge  OPT - \lambda^* \]

Let $a = \frac{1-Q_2}{Q_1-Q_2}$, then taking the convex combination of the above inequalities, we obtain:
$$ a  R(\lambda^*_-)  + (1-a)  R(\lambda^*_+)  \ge OPT$$
$$a Q(\lambda_-^*) + (1-a) Q(\lambda_+^*) = 1$$
This  completes the proof.
 \end{proof} 

To prove Theorem~\ref{thm:whittle-gap}, consider $n$ i.i.d. arms with $n \beta \ll 1$, $\alpha = \beta/(n-1)$
  and $r=1$. Each arm is in state $g$ w.p. $1/n$, so that all arms are in
  state $b$ w.p. $1/e$ and the maximum possible reward of any feasible
  policy is $1-1/e$ even with complete information about the states of all
  arms.

We will show that  Whittle's LP has value $1-O(\sqrt{n \beta})$ for $n \beta \ll 1$.  Since the LP is symmetric w.r.t. the arms, it is easy to show (from Theorem~\ref{thm:combine}) that for each arm, it will construct the same convex combination of two single-arm policies. The first policy is of the form $\P(t-1)$, and the second is of the form $\P(t)$.  The constraint is that if these policies are executed independently, exactly one arm is played in expectation per step. Since $\P(t)$ has lower average reward and rate of play than $\P(t-1)$, we consider the sub-optimal LP solution that uses policy $\P(t)$ for each arm.

The policy $\P(t)$ always plays in state $g$, and in state $b$, waits $t$
steps before playing. The value $t$ is chosen so that the rate of play for
each arm is less than $1/n$, and $\P(t-1)$ has a rate of play larger than
$1/n$.  The rate of play of the single arm policy $\P(t)$ is given by the formula:
$Q(t) = \frac{\beta + v_t}{t \beta + v_t}$. Since this is $1/n$, we have
$v_t = \beta (t-n)/(n-1)$. The reward of each arm is $R(t) = \frac{v_t}{t
  \beta + v_t} = \frac{t-n}{n(t-1)}$, so that the objective of Whittle's
LP is $n R(t) = 1-\Theta(n/t)$.

Now, from $v_t = \beta (t-n)/(n-1)$, we obtain  $1-(1-\beta')^t = \beta'(t-n)$, where $\beta' = \alpha + \beta = \beta \frac{n}{n-1}$. This holds for $t =  \Theta(\sqrt{n/\beta})$ provided $n\beta \ll 1$. Plugging this value of  $t$ into the value $n R(t)$ of Whittle's LP completes the proof of Theorem~\ref{thm:whittle-gap}. 

\subsection{Proof of Lemma~\ref{lem:whit}}
Recall the notation from Section~\ref{app:gap2} and Definition~\ref{def:rq}. We first present the following structural lemma about the optimal single-arm policy $L_i(\lambda)$. Suppose this policy is of the form $\P_i(t_i(\lambda))$, where $t_i(\lambda) = \mbox{argmax}_{t\ge 1} F_i(\lambda,t)$.  

\begin{lemma}
\label{lem:mon2}
$t_i(\lambda)$ is monotonically non-decreasing in $\lambda$.
\end{lemma}
\begin{proof}
We have: $\frac{\partial F_i(\lambda,t)}{\partial \lambda} = - Q_i(t) = - \frac{v_{it} + \beta_i}{v_{it} + t  \beta_i}$. Since $Q_i(t)$ is a decreasing function of $t$, the above is an increasing function and always negative, which implies that for smaller $t$, the function $F_i(\lambda,t)$ decreases faster as $\lambda$ is increased. This implies that if $t_i(\lambda) =  \mbox{argmax}_{t\ge 1} F_i(\lambda,t)$, then for $\lambda' \ge \lambda$, the maximum of $F_i(\lambda',t)$ is attained for some $t_i(\lambda') \ge t_i(\lambda)$.
\end{proof}

Now note that when $\lambda = 0$, there is no penalty, so that the single-arm policy maximizes its reward by playing every step regardless of the state. Therefore, $\Pi_i(s,t) \ge 0$ for all states $(s,t)$\footnote{Note that this is true {\em only} for {\sc Feedback MAB} where the underlying 2-state process evolves regardless of the plays; the claim need not be true for {\sc Monotone} bandits defined in Section~\ref{sec:recover}, where even with penalty $\lambda = 0$, the arm may idle in certain states.}.

Suppose the arm is in state $(g,1)$. The immediate expected reward if played is $r_i (1-\beta_i)$. If the penalty $\lambda < r_i (1-\beta_i)$,  a policy that plays the arm and stops later has positive expected reward minus penalty. Therefore, for penalty $\lambda$, the optimal decision at state $(g,1)$ is "play", so that $\Pi_i(g,1) \ge r_i (1-\beta_i)$.  We now show that $\Pi_i(g,1) = r_i (1-\beta_i)$. Suppose the penalty is $\lambda > r_i (1-\beta_i)$. If played in state $(g,1)$, the immediate expected reward minus penalty is negative, and leads to the policy being in state $(g,1)$ or $(b,1)$. The best possible total reward minus penalty in the future is obtained by always playing in state $(g,1)$ and waiting as long as possible in state $(b,1)$ (since this maximizes the chance of going to state $g$ if played). Whenever the arm is played in state $b$ after $w$ steps, the probability of observing state $g$ is at most $\frac{\alpha_i}{\alpha_i+\beta_i}$. Consider two consecutive events of the policy when the last play was in state $(g,1)$ and the current observed state is $(b,1)$. Since the optimal such policy is ergodic, this interval would define a renewal period. In this period, the expected penalty is at least $\lambda \left(\frac{\alpha+\beta}{\alpha} + \frac{1}{\beta}\right)$, and the expected reward is  $\frac{r_i}{\beta_i}$. Therefore,  the next expected reward minus penalty in the renewal period is: 

{\small \[ \frac{r_i - \lambda}{\beta_i} - \lambda
  \frac{\alpha_i+\beta_i}{\alpha_i}  <  r_i \left(1 -  (1-\beta_i)
    \left(1+ \frac{\beta_i}{\alpha_i}\right) \right) = - r_i
  \frac{\beta_i}{\alpha_i} (1-\beta_i - \alpha_i)  < 0 \]
}

The last inequality follows since $\alpha_i+ \beta_i \le 1-\delta$ for a $\delta > 0$ specified as part of input. This implies that if $\lambda > r_i (1-\beta_i)$, the any policy that plays in state $(g,1)$  has negative net reward minus penalty, showing that ``not playing" is optimal. Therefore, $\Pi_i(g,1) = r_i (1-\beta_i)$.

Next assume that for penalty slightly less than $r_i (1-\beta_i)$, the policy decision is to ``play" in state $(b,t)$. Consider the smallest such $t$.  Since the policy also decides to play in $(g,1)$, consider the renewal period defined by two consecutive events where the policy when the last play was in state $(g,1)$ and the current observed state is $(b,1)$. The reward is $\frac{r_i}{\beta_i}$ and the penalty is $\lambda \left(\frac1{v_{it}} + \frac{1}{\beta_i}\right)$. Since $\lambda = r_i(1-\beta_i)$, and $v_{it}  < \frac{\alpha_i}{\alpha_i+\beta_i}$, the above analysis shows that the net expected reward minus penalty is negative in renewal period. Therefore, the decision in $(b,t)$ is to ``not play", so that $\Pi_i(b,t) \le r_i (1-\beta_i)$.

Finally, for any $\lambda < r_i(1-\beta_i)$, consider the smallest $t \ge 1$ so that the optimal decision in state $(b,t)$ is to ``play". If this is finite, the optimal policy for this $\lambda$ is precisely $L_i(\lambda) = \P_i(t_i(\lambda))$. From Lemma~\ref{lem:mon2}, the function $t_i(\lambda)$ is non-decreasing in $\lambda$. Therefore, for any state $(b,t^*)$, the quantity $ \max\{ \lambda | L_i(\lambda) = \P_i(t^*)\}$ is well-defined. For larger values of penalty $\lambda$, we have $t_i(\lambda) > t^*$, so that the decision in $(b,t^*)$ is ``do not play". Therefore, $\Pi_i(b,t) = \max\{ \lambda | L_i(\lambda) = \P_i(t)\}$. Since $t_i(\lambda)$ is non-decreasing in $\lambda$, the function $\Pi_i(b,t)$ is non-decreasing in $t$. This completes the proof of Lemma~\ref{lem:whit}.

\end{document}